\documentclass{osa-article}

\journal{oe}
\articletype{Research Article}
\usepackage{algorithm,algorithmic,minibox,multirow,hhline}
\usepackage{float,verbatim,amsmath,amsfonts,bm,comment,enumerate,}
\usepackage{xcolor,makecell}

\begin{document}
	
\title{Parameter retrieval methods in ptychography}
	
	\author{Xukang Wei,\authormark{1,*} H. Paul Urbach\authormark{1}, Peter van der Walle,\authormark{2} and Wim M. J. Coene\authormark{1,3}}
	
	\address{\authormark{1}Optics Research Group, Imaging Physics Department, Delft University of Technology, The Netherlands\\
    \authormark{2}TNO Optics Department, Delft, The Netherlands\\	\authormark{3}ASML Netherlands B.V, Eindhoven, The Netherlands}
	
	\email{\authormark{*}x.wei-2@tudelft.nl} %% email address is required
	
	% \homepage{http:...} %% author's URL, if desired
	
	%%%%%%%%%%%%%%%%%%% abstract %%%%%%%%%%%%%%%%
	%% [use \begin{abstract*}...\end{abstract*} if exempt from copyright]
	
\begin{abstract}
We present a parameter retrieval method which incorporates prior knowledge about the object into ptychography. The proposed method is applied to two applications: (1) parameter retrieval of small particles from Fourier ptychographic dark field measurements; (2) parameter retrieval of a rectangular structure with real-space ptychography. The influence of Poisson noise is discussed in the second part of the paper. The Cram\'{e}r Rao Lower Bound in both applications is computed and Monte Carlo analysis is used to verify the calculated lower bound. With the computation results we report the lower bound for various noise levels and analyze the correlation of particles in application 1. For application 2 the correlation of parameters of the rectangular structure is discussed.
\end{abstract}
	
%%%%%%%%%%%%%%%%%%%%%%%%%%  body  %%%%%%%%%%%%%%%%%%%%%%%%%%
\section{Introduction}
Ptychography\cite{Hoppe1969,Rodenburg1992a,Chapman1996,Faulkner2004,Rodenburg2004,Guizar-Sicairos2008} is a scanning coherent diffraction imaging method for reconstructing a complex valued object function from intensity measurements recorded in the Fraunhofer or Fresnel diffraction region. In ptychography the object is partially illuminated multiple times with varying position of the illumination spot, so that the entire object is covered and adjacent illuminations partially overlap \cite{Silva2015}. The technique provides a solution to the so-called 'phase problem' and is found to be very suitable for EUV \cite{Seaberg2014,Odstrcil2015} and X-ray imaging applications \cite{Rodenburg2007,Thibault2008,Chapman2010,Pfeiffer2017} due to its high fidelity and its minimum requirement on optical imaging elements. Moreover, abundant studies show that ptychography is able to provide a wide field-of-view and retrieve the illumination probe also \cite{Thibault2009,Maiden2009}. During the last two decades, ptychography has been successfully demonstrated with X-ray radiation sources \cite{Thibault2008,Holler2017,Gardner2017}, electron beams \cite{Jiang2018} and visible light sources \cite{Maiden2017}.

More recently, Fourier ptychographic microscopy \cite{Zheng2013,Ou2014} has been proposed, which can be regarded as an extension of ptychography\cite{Horstmeyer2014}. The technique overcomes the resolution limit of conventional microscopy by enlarging the effective cut-off spatial frequency in the pupil plane. This is done by applying several plane wave illuminations to the sample. The detector is in the image conjugate to the sample plane, and each measurement corresponds to a particular incident angle of the illumination. With each tilted illumination, the diffraction pattern of the sample is shifted in the plane of the exit pupil of the lens, over the aperture used for imaging. Consecutive illumination tilts generate partially overlapping diffraction patterns within the aperture. With all of the Fourier ptychographic measurements, the spatial spectrum of the sample can be synthesized by using ptychographic algorithms with interchanged real space and reciprocal space coordinates \cite{Yeh2015,Horstmeyer2015,Zhang2015}. 

In general, the framework of real-space ptychograghy can be described as follows.  Let $\textbf{r}$ and $\textbf{k}$ be 3D coordinates in real space and reciprocal space:
\begin{align}
\textbf{r}\,=\,\left[x,y,z\right]^{T}\,=\,\left[\textbf{r}_{\perp},z\right]^{T},\qquad\textbf{k}\,=\,\left[k_{x},k_{y},k_{z}\right]^{T}\,=\,\left[\textbf{k}_{\perp},k_{z}\right]^{T}.
\label{eq.4-1-1}
\end{align} 
and $O(\textbf{r}_{\perp})$ the object transmission function. We use a laterally shifted probe, denoted by $P(\textbf{r}_{\perp})$, to illuminate the object multiple times. For the $j$th illumination, the exit wave immediately behind the object is: 
\begin{align}
\varPsi_{j}(\textbf{r}_{\perp})\,&=\,P(\textbf{r}_{\perp}-\textbf{R}_{\perp,j})\cdot O(\textbf{r}_{\perp})\,=\,P_{j}(\textbf{r}_{\perp})\cdot O(\textbf{r}_{\perp}),\label{eq.4-1-2}
\end{align} 
where $\textbf{R}_{\perp,j}$ specifies the shift of the $j$th illumination.
The probe function is assumed to have a finite support with, for instance, a circular boundary:
\begin{eqnarray}
P(\mathbf{r}_{\perp})\,=\,\left\lbrace 
\begin{tabular}{ll}
$P(\mathbf{r}_{\perp}),$ & $\left|\mathbf{r}_{\perp}\right|\leq r_{0},$\\
$0,$ & $\left|\mathbf{r}_{\perp}\right|>r_{0}.$
\end{tabular}
\right.\label{eq.4-1-3}
\end{eqnarray}
For a detector located at distance $z$ in the far field, the diffraction intensity pattern $I(\textbf{r}')$ for the $j$th illumination is:
\begin{align}
I_{j}(\textbf{r}'_{\perp})\,&=\,\left|\iint \varPsi_{j}(\textbf{r}_{\perp})\exp\left(-\text{i}\frac{2\pi}{\lambda z}\textbf{r}_{\perp}\cdot\textbf{r}'_{\perp}\right)d\textbf{r}_{\perp}\right|^{2}\,=\,\left|\mathcal{F}\left(\varPsi_{j}\right)\left(\textbf{k}'_{\perp}\right)\right|^{2}.\label{eq.4-1-4}
\end{align}
where $\mathcal{F}$ is the Fourier transform operator.  $\textbf{r}'_{\perp}$ is 2D coordinate in the detector plane. The relation between $\textbf{r}'_{\perp}$ and $\textbf{k}'_{\perp}$ is: $\textbf{k}'_{\perp}\,=\,2\pi\textbf{r}'_{\perp}(\lambda z)^{-1}$.

The task of ptychography is to find an estimate of the object which fits the given \textit{a priori} knowledge, while a cost function $\mathcal{E}$ is minimized. For the case of real-space ptychography, the \textit{a priori} knowledge is the exact information of the probe function and the set of relative positions $\mathbf{R}_{j}$.
The cost function $\mathcal{E}$ is defined as the $l_{2}$-distance between the modulus of the far field diffraction pattern $\left|\mathcal{F}\left(\varPsi_{j}\right)(\textbf{k}'_{\perp})\right|$ and the square root of the measured intensity $I_{j}^{\text{m}}(\textbf{k}'_{\perp})$:
\begin{align}
\mathcal{E}\,=\,\sum_{j}\mathcal{E}_{j}\,=\,\sum_{j}\sum^{N_{x}^{\text{det}},N_{y}^{\text{det}}}_{\textbf{k}'_{\perp}}\left[\sqrt{I_{j}^{\text{m}}(\textbf{k}'_{\perp})}-\left|\mathcal{F}\left(\varPsi_{j}\right)(\textbf{k}'_{\perp})\right|\right]^{2},\label{eq.4-1-5}
\end{align}
where $N^{\text{det}}_{x}$ and $N^{\text{det}}_{y}$ are the number of pixels of the detector in $x$-axis and $y$-axis, respectively. One way to minimize $\mathcal{E}$ is to use the gradient descent method. If we apply the gradient descent method to each $\mathcal{E}_{j}$ sequentially, the algorithm is equivalent to the ptychography iterative engine (PIE)\cite{Rodenburg2004,Guizar-Sicairos2008}. Another popular choice is the difference map algorithm, which is formulated in terms of finding the intersection of two constraint sets\cite{Elser2003,Thibault2009}. When the ptychographic measurements contain a relatively large amount of noise, one can utilize de-noising ptychographic algorithms to obtain a better image of the object. One of the most powerful and robust de-noising methods is the Maximum Likelihood estimation\cite{Godard2012,Thibault2012,Yeh2015,Odstrcil2018}, which requires knowledge of the noise model. 

The ptychographic measurement $I_{j}(\textbf{k}'_{\perp})$ is commonly recorded by a 2D detector, e.g. a charge-coupled device (CCD). Therefore $\textbf{k}'_{\perp}$ is a discretized grid and is meshed according to the distance $z$ and the size of pixel of the detector. The retrieved object function, denoted by $\hat{O}$, is also on a discretized grid $\textbf{r}_{\perp}$. $\textbf{r}_{\perp}$ and $\textbf{k}'_{\perp}$ have the relation:
\begin{align}
\left[\Delta x,\Delta y\right]^{T}\,=\,2\pi\left[(N_{x}^{\text{det}}\Delta k'_{x})^{-1},(N_{y}^{\text{det}}\Delta k'_{y})^{-1}\right]^{T},\label{eq.4-1-6}
\end{align}
where $\Delta x$ and $\Delta y$ are the spacing of a single grid cell in $x$-axis and $y$-axis, respectively, and $\Delta k'_{x}$ and $\Delta k'_{y}$ are the spacing of a grid cell in $k_{x}$ and $k_{y}$, respectively. Note that the total field-of-view (FoV) in the object plane is:
\begin{align}
\text{FoV}\,=\,\left[N_{x}\Delta x,N_{x}\Delta y\right]^{T},\label{eq.4-1-7}
\end{align}
where $N_{x}>N_{x}^{\text{det}}$ and $N_{y}>N_{y}^{\text{det}}$ due to that ptychography is a scanning imaging technique which provides an extended FoV. In line with this extended FoV, we have the effective spacing of the grid cell in the reciprocal space:
\begin{align}
\left[\Delta k_{x},\Delta k_{y}\right]^{T}\,=\,\left[(N_{x})^{-1}N_{x}^{\text{det}}\Delta k'_{x},(N_{y})^{-1}N_{y}^{\text{det}}\Delta k'_{y}\right]^{T}.\label{eq.4-1-8}
\end{align}

We can see that, when the influence of noise is negligible, the relation given in Eq. (\ref{eq.4-1-6}) imposes a resolution limit to the reconstruction in ptychography. To overcome this limit, several 'superresolution' methods have been proposed \cite{Maiden2011,Szameit2012,Sidorenko2015}. One of the ideas lying behind these methods is to impose additional \textit{a priori} knowledge about the object, e.g. analytic continuation of the Fourier transform of bounded support\cite{Slepian1961,Papoulis1975,Delsarte1985} or sparsity\cite{Szameit2012,Sidorenko2015}, to the algorithm. In this paper we show a parameter retrieval algorithm which incorporates additional \textit{a priori} knowledge about the object into ptychography. We present this algorithm by numerically demonstrating two applications: 
\begin{enumerate}[(1)]
	\item Parameter retrieval of sub-wavelength particles using Fourier ptychography with dark field measurements only. For this example the configuration is in line with the 'RapidNano' particle scanner developed by TNO \cite{Walle2014,Walle2017}. The particle scanner is supposed to detect nano-particles on an EUV reticle. Since only dark field images are recorded in the scanner, the part of the spatial spectrum of the object in the neighborhood of $\left|\textbf{k}_{\perp}\right|=0$ is lost. The missing data can in principle be filled in by analytic continuation using the fact that the object has bounded support, however, this method is unstable with noisy measurement and leads in practice to incorrect reconstructions\cite{Delsarte1985,Ferreira1994}. However. as shown in Section 2, the proposed parameter retrieval algorithm is able to extract information of sub-wavelength particles from the incomplete data.
	\item Parameter retrieval of rectangular objects using real-space ptychography. This example can also be applied to the metrology of EUV reticles \cite{Boef2016,Ansuinelli2019}. We demonstrate the proposed parameter retrieval method for this application in Section 3.
\end{enumerate}

To study the influence of Poisson noise on the proposed parameter retrieval scheme, we compute the Cram\'{e}r Rao Lower Bound (CRLB) and perform Monte Carlo analysis for both two applications in the second part of this paper. We derive the general form of the Fisher information matrix in Section 4. For application 1, the calulated CRLB and Monte Carlo result are shown in Section 5. For application 2, the discussion about the correlation of the parameters of the rectangular structure can be found in Section 6.

\section{Application 1: parameter retrieval of sub-wavelength particles using Fourier ptychography with dark field measurement}

\subsection{Description of the 'RapidNano 3' particle scanner}
The 'RapidNano 3' particle scanner \cite{Walle2014,Walle2017} is designed to detect small dielectric particles on a flat substrate. The particles are made of polystyrene latex (PSL) beads and the typical diameter of the particle is $\sim50nm$. The scanner has detection limit of 42 nm PSL particles, i.e. the capture rate is 95\% at this size. Note that the particles on the substrate can be any material and PSL is only the calibration standard. The particles are sparsely distributed on the sample mostly. The substrate is reflective, made of silicon, and its lateral size can be up to 6x6 inch, i.e. the size of an EUV mask. The illumination is a $532nm$, $p$-polarized, fully coherent plane wave laser beam. The incident angle of the illumination is $60$ degree, with $9$ regularly distributed azimuth incident directions around $360$ degree. The NA of the objective lens is 0.4, therefore the measurement is a dark field image of the sample as is illustrated in Fig. \ref{Fig.4-4}. 

\subsection{Single dipole radiation}
Considering that the diameter of the detected particles is around $10$ times smaller than the illumination wavelength, we begin by using the dipole radiation formula to model the wavefield scattered by the particles. Suppose that there are $N$ dipoles in the plane $z=0$, and the $i$th oscillating dipole is located at position $\textbf{r}_{i}=\left[\textbf{r}_{\perp,i},0\right]^{T},i=1,2,\cdots,N$, and is excited by an incident plane wave $\textbf{E}_{\text{in},j}$:
\begin{align}
\textbf{E}_{\text{in},j}\,=\,A_{\text{in}}e^{\text{i}\textbf{k}_{j}\cdot\textbf{r}}\hat{\textbf{e}}_{p}(\textbf{k}_{j})\,=\,A_{\text{in}}e^{\text{i}\textbf{k}_{\perp,j}\cdot\textbf{r}_{\perp}}\hat{\textbf{e}}_{p}(\textbf{k}_{j}),\label{eq.4-2-1}
\end{align}
where $A_{\text{in}}^{2}$ is proportional to the illumination power and $\hat{\textbf{e}}_{p}(\textbf{k}_{j})$ denotes the polarization direction. 

\begin{comment}
Since we have $p$-polarized incident electric wave in the particle scanner, $\hat{\textbf{e}}_{p}(\textbf{k}_{j})$ is given by:
\begin{align}
\hat{\textbf{e}}_{p}(\textbf{k}_{j})\,=\,\frac{1}{k\left|\textbf{k}_{\perp,j}\right|}\left[
\begin{tabular}{c}
$k_{x,j}k_{z,j}$\\
$k_{y,j}k_{z,j}$\\
$-\left|\textbf{k}_{\perp,j}\right|^{2}$
\end{tabular}
\right].\label{eq.4-2-3}
\end{align}
\end{comment}

For the $i$th dipole with position $\textbf{r}_{\perp,i}$, we denote the dipole moment by:
\begin{align}
\textbf{p}_{i,j}\,=\,{\alpha}_{i}\textbf{E}_{\text{in},j}\,=\,{\alpha}_{i}A_{\text{in}}e^{\text{i}\textbf{k}_{\perp,j}\cdot\textbf{r}_{\perp,i}}\hat{\textbf{e}}_{p}(\textbf{k}_{j}),\label{eq.4-2-2}
\end{align}
where $\epsilon_{0}$ is the permittivity of free space and ${\alpha}_{i}$ is the polarisability of the particle. For a dielectric sphere with diameter $d$, the dipole moment $\textbf{p}_{i,j}^{\text{sphere}}$ in the quasi-static approximation is given by:
\begin{align}
\textbf{p}_{i,j}^{\text{sphere}}\,&=\,\left(\frac{\epsilon_{r}-2}{\epsilon_{r}+1}\right){d}_{i}^{3}\textbf{E}_{\text{in},j},\label{eq.4-2-3}
\end{align}
where $\epsilon_{r}=n^{2}_{\text{PSL}}$ is the relative permittivity of the dielectric. $n_{\text{PSL}}$ is the refractive index of the small particles. Since the real part of $n_{\text{PSL}}$ is $\sim10^{6}$ times larger than the imaginary part, i.e. than the absorption index, we assume the ${\alpha}_{i}$ is real valued for the rest of this paper. We see that ${\alpha}_{i}$ is proportional to the volume of the dielectric particle. 

The electric field radiating from the $i$th dipole due to the $j$th illumination is given by\cite{Griffiths1999,Novotny2012}:
\begin{align}
\textbf{E}_{\text{scat},i,j}\,&=\,\stackrel{\leftrightarrow}{\textbf{G}}\left(\textbf{r},\textbf{r}_{i}\right)\textbf{p}_{i,j},\label{eq.4-2-4}
\end{align}
where $\stackrel{\leftrightarrow}{\textbf{G}}\left(\textbf{r},\textbf{r}_{i}\right)$ is the dyadic Green’s function:
\begin{align}
\stackrel{\leftrightarrow}{\textbf{G}}\left(\textbf{r},\textbf{r}_{i}\right)\,&=\,\frac{k^{2}}{4\pi\epsilon_{0}}\left(\stackrel{\leftrightarrow}{\textbf{I}}+\frac{1}{k^{2}}\nabla\nabla\right)\frac{e^{\text{i}k\left|\textbf{r}-\textbf{r}_{i}\right|}}{\left|\textbf{r}-\textbf{r}_{i}\right|},\label{eq.4-2-14}
\end{align}
where $\stackrel{\leftrightarrow}{\textbf{I}}$ is the $3\times3$ identity matrix. Considering that the detector of the particle scanner is insensitive to the polarization state and that the NA of the objective lens is 0.4, we ignore the effect of the polarization of the wavefield for simplicity. Hence we arrive at a scalar scattered amplitude given by:
\begin{align}
E_{\text{scat},i,j}\,&=\,A_{\text{in}}k^{2}{\alpha}_{i}e^{\text{i}\textbf{k}_{\perp,j}\cdot\textbf{r}_{\perp,i}}G\left(\textbf{r},\textbf{r}_{i}\right),\label{eq.4-2-5}
\end{align}
where
\begin{align}
G\left(\textbf{r},\textbf{r}_{i}\right)\,&=\,\frac{k^{2}}{4\pi\epsilon_{0}}\frac{e^{\text{i}k\left|\textbf{r}-\textbf{r}_{i}\right|}}{\left|\textbf{r}-\textbf{r}_{i}\right|}.\label{eq.4-2-15}
\end{align}

\begin{comment}
\begin{align}
\tilde{\textbf{E}}_{\text{scat},i,j}\,&=\,Ak^{2}\alpha_{i}e^{\text{i}\textbf{k}_{\perp,j}\cdot\textbf{r}_{\perp,i}}\frac{e^{\text{i}k\left|\textbf{R}_{1}\right|}}{\left|\textbf{R}_{1}\right|}
\left[ 
\begin{tabular}{ccc}
$1$&$0$&$0$\\
$0$&$1$&$0$\\
$0$&$0$&$0$
\end{tabular}
\right]
\hat{\textbf{e}}_{p}(\textbf{k}_{j})\nonumber\\
&=\,Ak^{2}\alpha_{i}e^{\text{i}\textbf{k}_{\perp,j}\cdot\textbf{r}_{\perp,i}}\frac{e^{\text{i}k\left|\textbf{R}_{1}\right|}}{\left|\textbf{R}_{1}\right|}\textbf{s}_{j},\label{eq.4-1-10}
\end{align}
where
\begin{align}
\textbf{s}_{j}\,=\,\frac{k_{z,j}}{k\left|\textbf{k}_{\perp,j}\right|}\left[k_{x,j},k_{y,j},0\right]^{T}.\label{eq.4-1-11}
\end{align}
\end{comment}

\subsection{Dark field measurement from the particle scanner}
By Fourier transforming Eq. (\ref{eq.4-2-5}) with respect to $\textbf{r}_{\perp}$, we have:
\begin{align}
\mathcal{F}\left(E_{\text{scat},i,j}\right)(\textbf{k}_{\perp},z)\,&=\,A_{\text{in}}k^{2}\frac{e^{\text{i}k_{z}\left|z\right|}}{8\text{i}\pi\epsilon_{0}k_{z}}{\alpha}_{i}e^{-\text{i}\textbf{r}_{\perp,i}\cdot\left(\textbf{k}_{\perp}-\textbf{k}_{\perp,j}\right)}.\label{eq.4-2-6}
\end{align}
$\mathcal{F}\left(E_{\text{scat},i,j}\right)$ can be regarded as the 2D spatial spectrum of the scattered wavefield in the plane $z$. The electric field at the exit pupil is given by:
\begin{align}
\mathcal{F}\left(E_{\text{scat},i,j}\right)_{\text{exit}}(\textbf{k}_{\perp},z)\,&=\,\textbf{1}_{k\text{NA}}(\textbf{k}_{\perp})A_{\text{in}}k^{2}\frac{e^{\text{i}k_{z}\left|z\right|}}{8\text{i}\pi\epsilon_{0}k_{z}}{\alpha}_{i}e^{-\text{i}\textbf{r}_{\perp,i}\cdot\left(\textbf{k}_{\perp}-\textbf{k}_{\perp,j}\right)},\label{eq.4-2-7}
\end{align}
where $\textbf{1}_{k\text{NA}}(\textbf{k}_{\perp})$ represents the numerical aperture of the objective lens:
\begin{align}
\textbf{1}_{k\text{NA}}(\textbf{k}_{\perp})\,&=\,
\left\lbrace 
\begin{tabular}{cc}
$1,$&$\left|\textbf{k}_{\perp}\right|\leq k\text{NA},$\\
$0,$&$\left|\textbf{k}_{\perp}\right|> k\text{NA}.$\\
\end{tabular}
\right.\label{eq.4-2-8}
\end{align}

By summing over all the dipoles, we find the total field in the exit pupil, which is denoted by $\varPsi_{j}$:
\begin{align}
\varPsi_{j}(\textbf{k}_{\perp},z)\,&=\,\sum_{i}\mathcal{F}\left(E_{\text{scat},i,j}\right)_{\text{exit}}(\textbf{k}_{\perp},z)\nonumber\\
&=\,\textbf{1}_{k\text{NA}}(\textbf{k}_{\perp})A_{\text{in}}k^{2}\frac{e^{\text{i}k_{z}\left|z\right|}}{8\text{i}\pi\epsilon_{0}k_{z}}\sum_{i}{\alpha}_{i}e^{-\text{i}\textbf{r}_{\perp,i}\cdot\left(\textbf{k}_{\perp}-\textbf{k}_{\perp,j}\right)}\nonumber\\
&=\,Q(\textbf{k}_{\perp},z)\cdot O(\textbf{k}_{\perp}-\textbf{k}_{\perp,j}),\label{eq.4-2-9}
\end{align}
where
\begin{align}
&Q(\textbf{k}_{\perp},z)\,=\,\textbf{1}_{k\text{NA}}(\textbf{k}_{\perp})A_{\text{in}}k^{2}\frac{e^{\text{i}k_{z}\left|z\right|}}{8\text{i}\pi\epsilon_{0}k_{z}},\label{eq.4-2-10}
\end{align}
and $O(\textbf{k}_{\perp})$ is the Fourier transform of the object defined by
\begin{align}
&O(\textbf{k}_{\perp})\,=\,\sum_{i}\alpha_{i}e^{-\text{i}\textbf{k}_{\perp}\cdot\textbf{r}_{\perp,i}}.\label{eq.4-2-17}
\end{align}
Note that the object function is assumed to be independent of the angle of incidence, i.e. the only effect of the tilted illumination is the shift of the Fourier transform of the object function over the pupil plane.
Finally, by inverse Fourier transforming $\varPsi_{j}$ and taking the squared modulus, we arrive at the expression for the measured intensity in the detector plane:
\begin{align}
I_{j}(\textbf{r}_{\perp}',z)\,&=\,\left|\mathcal{F}^{-1}\left(\Psi_{j}\right)\right|^{2}(\textbf{r}_{\perp}',z)\nonumber\\
&=\,\left|\mathcal{F}^{-1}\left[Q(\textbf{k}_{\perp}+\textbf{k}_{\perp,j},z)\cdot O(\textbf{k}_{\perp})\right]\right|^{2}(\textbf{r}_{\perp}',z),\label{eq.4-2-11}
\end{align}
where $\textbf{r}_{\perp}'$ is the 2D regular grid.

\begin{figure}[htp!]
	\centering\includegraphics[width=0.75\textwidth]{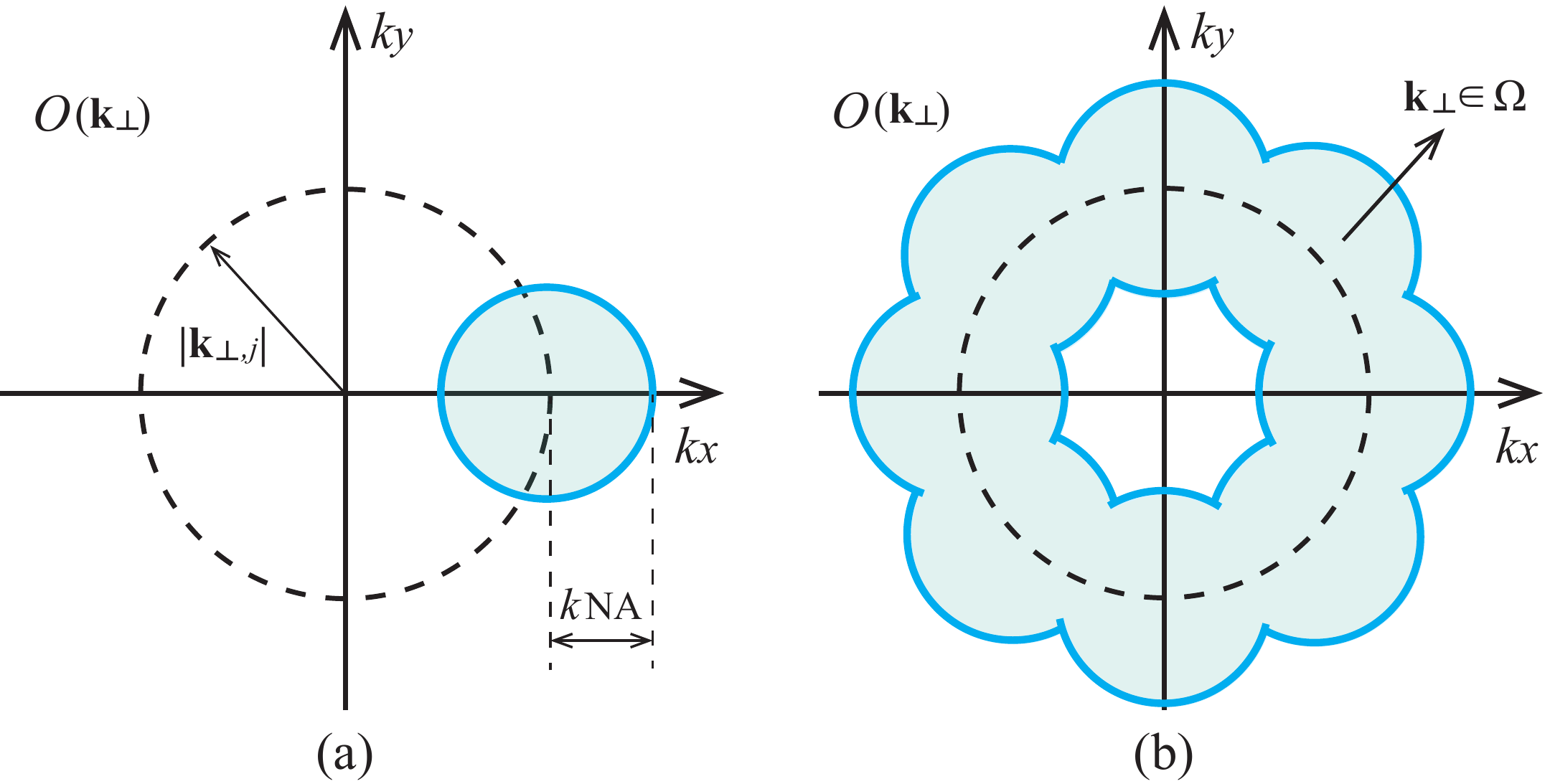}	\caption{Graphical illustration of $O(\textbf{k}_{\perp})$. (a) The blue disk is defined by $\textbf{1}_{k\text{NA}}(\textbf{k}_{\perp})$ and indicates information about $O$ included in the single measurement $I_{j}(\textbf{r}_{\perp}',z)$. (b) The retrievable part of $O$ from all given dark field measurements.}\label{Fig.4-4}
\end{figure}
For the configuration of the particle scanner, $\left|\textbf{k}_{\perp,j}\right|$ is fixed and equal to $k\sin(\frac{\pi}{3})$. The NA of the objective lens is $\sim 0.4$. Therefore, the intensity measurements do not contain any information about $O(\textbf{k}_{\perp}=0)$ and its surrounding region, as shown in Fig. \ref{Fig.4-4}. The blue shaded area in Fig. \ref{Fig.4-4}(a) illustrates the information about $O(\textbf{k}_{\perp})$ included in the single measurement $I_{j}(\textbf{r}_{\perp}',z)$, while the blue shaded area in Fig. \ref{Fig.4-4}(b) represents the retrievable information from all measurements. We denote this retrievable part of $O$ by $\textbf{1}_{\Omega}O(\textbf{k}_{\perp})$:
\begin{align}
\textbf{1}_{\Omega}(\textbf{k}_{\perp})\,&=\,
\left\lbrace 
\begin{tabular}{cc}
$1,$&$\textbf{k}_{\perp}\in\Omega,$\\
$0,$&$\textbf{k}_{\perp}\notin\Omega,$\\
\end{tabular}
\right.\label{eq.4-2-16}
\end{align}
where $\Omega$ is the blue shaded region in Fig. \ref{Fig.4-4}(b).

\subsection{Retrieving the parameters of the particles}
To retrieve $\alpha_{i}$ and the position $\textbf{r}_{\perp,i}$ of the dipoles, we first reconstruct the complex valued function $\textbf{1}_{\Omega}O(\textbf{k}_{\perp})$ in the pupil plane from the set of intensity measurements $I_{j}(\textbf{r}_{\perp}')$. This can be done by applying a ptychographic algorithm. We use $\textbf{1}_{\Omega}\hat{O}(\textbf{k}_{\perp})$ to denote the reconstruction obtained by the ptychographic method.

Once $\textbf{1}_{\Omega}\hat{O}(\textbf{k}_{\perp})$ is obtained, we apply the method of least square to estimate $\alpha_{i}$ and $\textbf{r}_{\perp,i}$ of all dipoles. The number of freedom in this problem is $N\times3$, where $N$ is the number of dipoles within the field-of-view (FoV). Note that if ${\alpha}_{i}$ is complex valued, the degrees of freedom should be $N\times4$. When $N$ is in the order of $10^0\sim10^1$, we have much less degrees of freedom than in the traditional Fourier ptychography problem.

Our proposed parameter retrieval algorithm is shown in the following.
\begin{enumerate}[(1)]
\item Use a ptychographic algorithm to retrieve the complex valued wavefield $\textbf{1}_{\Omega}O(\textbf{k}_{\perp})$ in the pupil plane.
\item From all the dark field intensity measurements, find the lower and upper bound of $\alpha_{i}$ and $\textbf{r}_{\perp,i}$ for $i=1,2,\cdots,N$. These bounds are denoted by: $\alpha_{i}^{l}$, $\alpha_{i}^{u}$, $\textbf{r}_{\perp,i}^{l}$ and $\textbf{r}_{\perp,i}^{u}$.
\item Solve the following problem:
\begin{align}
\arg\min_{\alpha_{i},\textbf{r}_{\perp,i}}&\hspace{0.5em} \left\|\textbf{1}_{\Omega}\hat{O}(\textbf{k}_{\perp})-\sum_{i}\alpha_{i}e^{-\text{i}\textbf{k}_{\perp}\cdot\textbf{r}_{\perp,i}}\right\|_{\textbf{k}_{\perp}\in\Omega}^{2},\nonumber\\
\text{subject to}&\hspace{0.5em} \alpha_{i}^{l}\leq\alpha_{i}\leq\alpha_{i}^{u},\hspace{2.1em}i=1,2,\cdots,N, \nonumber\\
&\hspace{0.5em}\textbf{r}_{\perp,i}^{l}\preceq\textbf{r}_{\perp,i}\preceq\textbf{r}_{\perp,i}^{u},\hspace{0.5em}i=1,2,\cdots,N,\label{eq.4-2-12}
\end{align}
where $\preceq$ denotes vector inequality: $\textbf{r}_{\perp}^{l}\preceq\textbf{r}_{\perp}^{u}$ means $x^{l}\leq x^{u}$ and $y^{l}\leq y^{u}$.
\end{enumerate}

\subsection{Simulation}
\begin{figure}[htp!]
	\centering\includegraphics[width=0.6\textwidth]{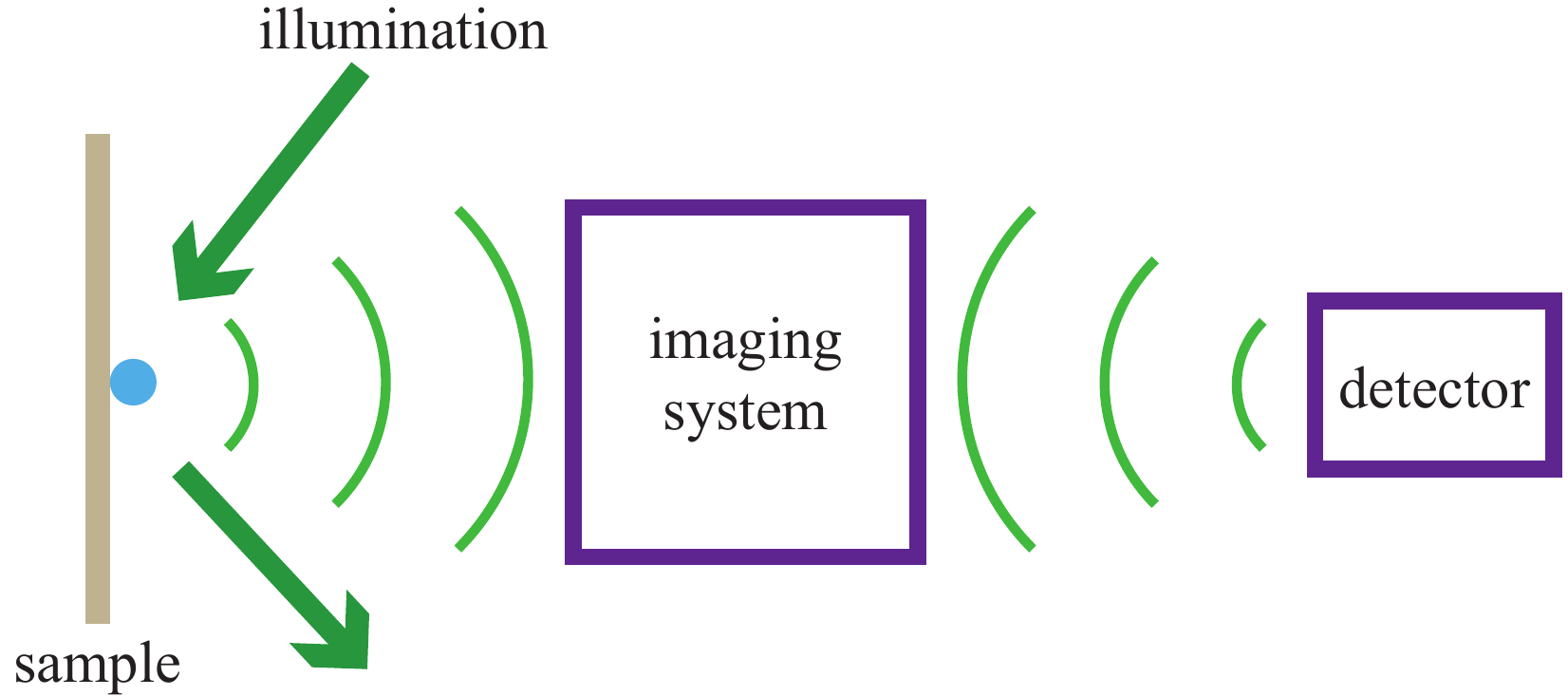}	
	\caption{Illustration of the setup of application 1. The incident angle of the illumination is 60 degree, with multiple azimuth incident directions around 360 degree.}\label{Fig.4-5}
\end{figure}

To validate the proposed parameter retrieval algorithm, a preliminary simulation is reported in this section. The configuration is drawn in Fig. \ref{Fig.4-5} and the parameter settings of the setup is described in Table \ref{table.4-3}. Since the NA of the imaging system is smaller than $\left|\textbf{k}_{\perp,j}\right|$, the measurements at the detector plane are always dark field images. We assume that the detector is insensitive to the polarization state of the wavefield.
\begin{table}
	\centering
	\caption{Configuration settings in the simulation}
\begin{tabular}{ccccc}
	\hline\hline
	\multicolumn{2}{c}{illumination} && \multicolumn{2}{c}{imaging system} \\
	\cline{1-2} \cline{4-5}
	wavelength & incident angle && NA & magnification \\
	\cline{1-2} \cline{4-5}
	$500nm$ & $60$ degree && $0.4$ & $20$ \\
	\hline\hline
	\multicolumn{3}{c}{detector} && \multicolumn{1}{c}{grid spacing in object plane}\\
	\cline{1-3} \cline{5-5}
	pixel size & pixel number & FoV && $\Delta x$ and $\Delta y$\\
	\cline{1-3} \cline{5-5}
	$5\mu m$ & $200\times200$ & $50\mu m$ && $133.3nm$ \\
	\hline\hline
\end{tabular}
\label{table.4-3}
\end{table}

\begin{figure}[htp!]
	\centering\includegraphics[width=1\textwidth]{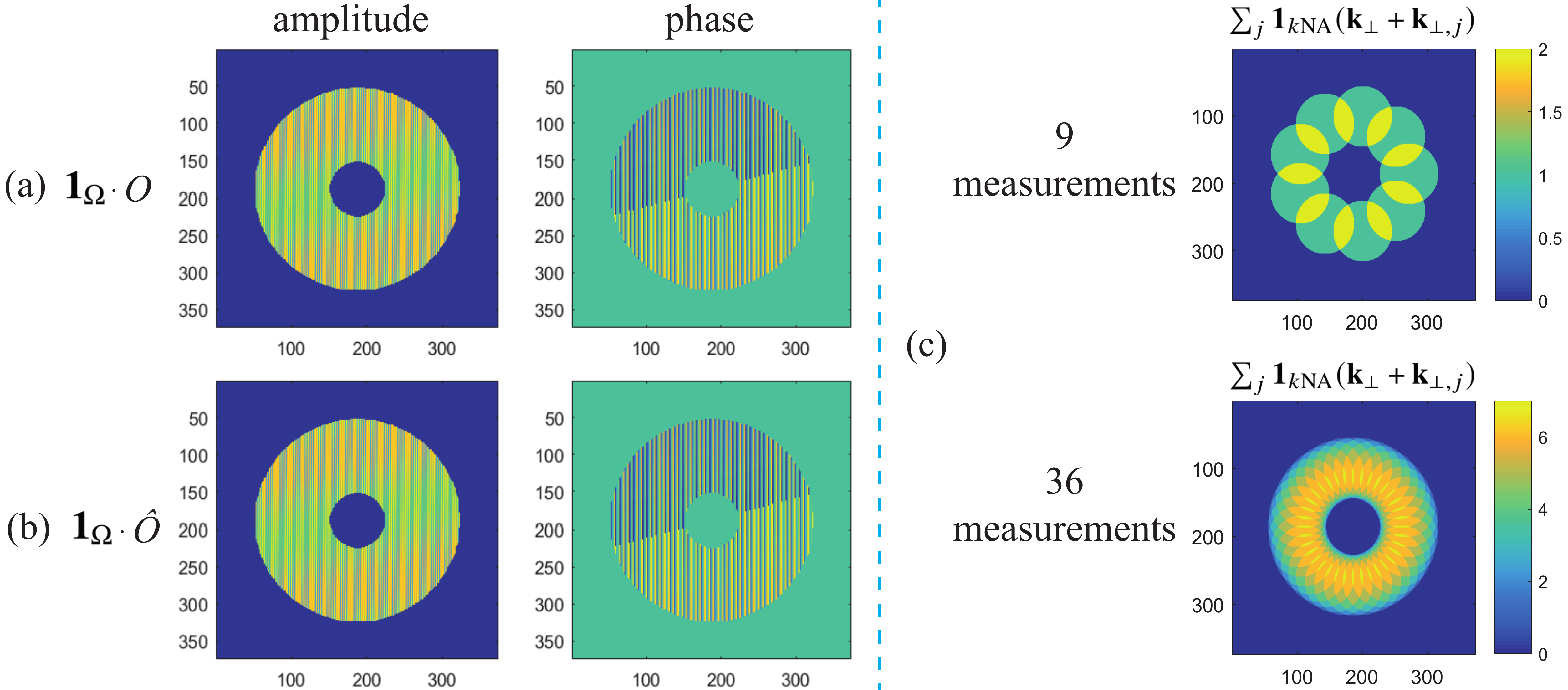}	
	\caption{(a) The amplitude and phase of the actual complex function $\textbf{1}_{\Omega}O(\textbf{k}_{\perp})$. (b) The amplitude and phase of $\textbf{1}_{\Omega}O(\textbf{k}_{\perp})$ which is reconstructed from the Fourier ptychographic algorithm. (c) Illustration of the illuminated area in the reciprocal space, i.e. $\sum_{j}\textbf{1}_{k\text{NA}}(\textbf{k}_{\perp}+\textbf{k}_{\perp,j})$, for 9 and 36 dark field measurements, respectively}\label{Fig.4-6}
\end{figure}

The simulated sample consists of two dipoles. The actual scattering strength $\alpha_{i}$ and the position $\textbf{r}_{\perp,i}$ of the dipoles are listed in Table \ref{table.4-4}. Based on these given parameters, we first construct the actual complex valued function $\textbf{1}_{\Omega}O(\textbf{k}_{\perp})$ according to Eq. (\ref{eq.4-2-10}). The dark field intensity measurements are noise-free and computed in accordance with Eq. (\ref{eq.4-2-11}). In line with the $1^{\text{st}}$ step of the proposed method given in Section 1.4, the reconstructed object function, denoted by  $\textbf{1}_{\Omega}\hat{O}(\textbf{k}_{\perp})$, is obtained by applying the Fourier ptychography method. We assume that the function $Q(\textbf{k}_{\perp}+\textbf{k}_{\perp,j},z)$ is known and we ignore the polarization state. In the simulation we notice that only $9$ incident plane waves cannot provide sufficient data redundancy. Instead we use $36$ plane waves with regularly distributed azimuth incident directions around $360$ degrees in this simulation. The actual function $O$ and the reconstructed one are shown in Fig. \ref{Fig.4-6}(a) and Fig. \ref{Fig.4-6}(b), respectively. Fig. \ref{Fig.4-6}(c) illustrates the illuminated area in the reciprocal space, i.e. $\sum_{j}\textbf{1}_{k\text{NA}}(\textbf{k}_{\perp}+\textbf{k}_{\perp,j})$, for 9 and 36 dark field measurements, respectively.
\begin{figure}[htp!]
	\centering\includegraphics[width=0.75\textwidth]{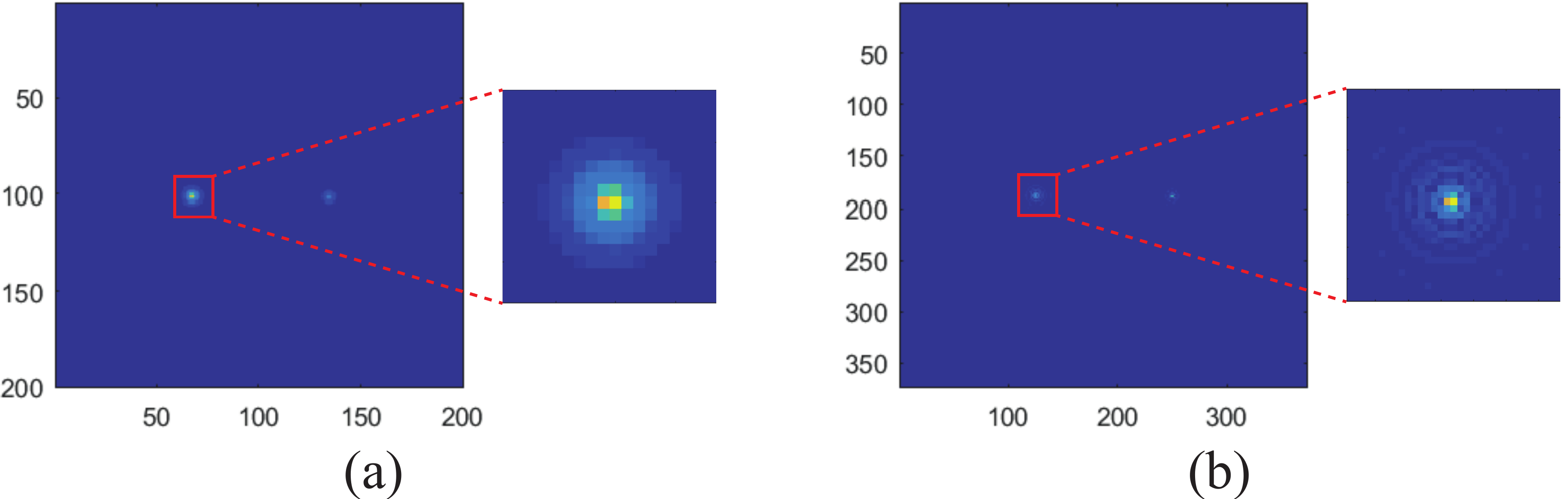}	
	\caption{(a) The incoherent sum of all 36 dark field measurements, i.e. $\sum_{j}I_{j}(\textbf{r}'_{\perp})$. (b) The amplitude of scattering wavefield at plane $z\rightarrow0$, i.e. $\left|\mathcal{F}^{-1}(\textbf{1}_{\Omega}\hat{O})\right|^{2}(\textbf{r}_{\perp})$, which is reconstructed with the Fourier ptychography method. The inserted graphs correspond to the dipole $i=1$.}\label{Fig.4-7}
\end{figure}

In Fig. \ref{Fig.4-7}(a) we show the incoherent sum of all 36 simulated noise-free intensity measurements, i.e. $\sum_{j}I_{j}(\textbf{r}'_{\perp})$, and in Fig. \ref{Fig.4-7}(b) we present the squared amplitude of the scattered field from the sample at plane $z\rightarrow0$, i.e. $\left|\mathcal{F}^{-1}(\textbf{1}_{\Omega}\hat{O})\right|^{2}(\textbf{r}_{\perp})$. For application 1 the spacing of grid  $\textbf{r}'_{\perp}$ and $\textbf{r}'_{\perp}$ fulfills:
\begin{align}
\left[\Delta x,\Delta y\right]^{T}\,=\,\left[(N_{x})^{-1}N_{x}^{\text{det}}\Delta x',(N_{y})^{-1}N_{y}^{\text{det}}\Delta y'\right]^{T},\label{eq.4-2-13}
\end{align}
which can be derived from Eq. (\ref{eq.4-1-8}) by interchanging the real space and reciprocal space coordinates. The inserted graphs in Fig. \ref{Fig.4-7} correspond to dipole $i=1$. In line with Table. \ref{table.4-3}, every dark field measurement is a $200\times200$ array with a $250nm$ pixel size. The reconstructed scattered field shown in Fig. \ref{Fig.4-7}(b) only contains information of $\textbf{k}_{\perp}\in\Omega$. The side-lobe which appears in the neighborhood of the particles in Fig. \ref{Fig.4-7}(b) is due to the fact that the reconstruction is convoluted by $\mathcal{F}^{-1}(\textbf{1}_{\Omega})(\textbf{r}_{\perp})$. Without knowing the wavefield at $\textbf{k}_{\perp}=0$ and its surrounding region or without considering any prior information about the sample, the reconstructed scattering field cannot provide a unique physical solution.

Once $\textbf{1}_{\Omega}\hat{O}(\textbf{k}_{\perp})$ is obtained, we retrieve $\alpha_{i}$ and $\textbf{r}_{\perp,i}$ by minimizing the least square function given in Eq. (\ref{eq.4-2-12}). This is done by using the 'fmincon' solver in MATLAB. To facilitate the solver to find the global minimum, a proper starting search point and a set of bounds for $\alpha_{i}$ and $\textbf{r}_{\perp,i}$ are needed. From Fig. \ref{Fig.4-7} we see that one can deduce a guess about the scattering strength and the position of the dipoles from the dark field measurements. Based on the guess we can obtain the starting point and the bounds. The accuracy of the guess of the position is limited by the pixel size of the detector. In the simulation we deduce the initial guess as follows. We first choose in Fig. \ref{Fig.4-7}(b) one pixel cell which approximately have minimal and equal distances from the centers of the images of two dipoles. In Fig. \ref{Fig.4-7}(b) the indices of this pixel cell in the $x$ and $y$ directions are $\left[195,186\right]^{T}$. Then we set the top left corner of this pixel cell as origin. The initial guess of position of the dipoles are obtained by roughly measuring the distance between the origin and the center of the image of the dipoles in Fig. \ref{Fig.4-7}(b). To determine the bounds of the position, we first choose two $5\times5$ pixel arrays which center at the brightest pixel cells of the image of two dipoles, respectively. The bounds of the positions are determined by the outer boundary of the two $5\times5$ pixel arrays in the $x$ and $y$, respectively. The initial guess of the the scattering strength of each dipole, on the other hand, is determined by summing the absolute square of the value on every grid cells over the image of each dipole. In the simulation we use a random number generator to create a starting search point which is close to the actual parameters. The starting point of all parameters are shown in Table. \ref{table.4-4}. The retrieved parameters are listed in the most right column of the same table.
\begin{table}
	\centering
	\caption{Retrieved parameters of two dipoles in the noise free simulation}
	\begin{tabular}{c||ccc}
		\hline
        & actual value & initial guess & retrieved value \\
		\hhline{=||===}
		$\alpha_{1}/(\lambda^{3})$ & $1.000\times10^{-3}$ & $1.195\times10^{-3}$ & $1.000\times10^{-3}$ \\
		\hline
		$x_{1}$ & $-8.333\,\mu m$ & $-8.349\,\mu m$ & $-8.333\,\mu m$ \\
		\hline
		$y_{1}$ & $0.000\,\mu m$ & $0.113\,\mu m$ & $0.000\,\mu m$ \\
		\hhline{=||===}
		$\alpha_{2}/(\lambda^{3})$ & $0.512\times10^{-3}$ & $0.329\times10^{-3}$ & $0.512\times10^{-3}$ \\
		\hline
		$x_{2}$ & $8.356\,\mu m$ & $8.327\,\mu m$ & $8.356\,\mu m$ \\
		\hline
		$y_{2}$ & $0.088\,\mu m$ & $-0.029\,\mu m$ & $0.088\,\mu m$ \\
		\hline
	\end{tabular}
	\label{table.4-4}
\end{table}

\section{Application 2: parameter retrieval of a rectangular object using real-space ptychography}
\subsection{Single object embedded in constant surrounding}
\begin{figure}[htp!]
	\centering\includegraphics[width=0.55\textwidth]{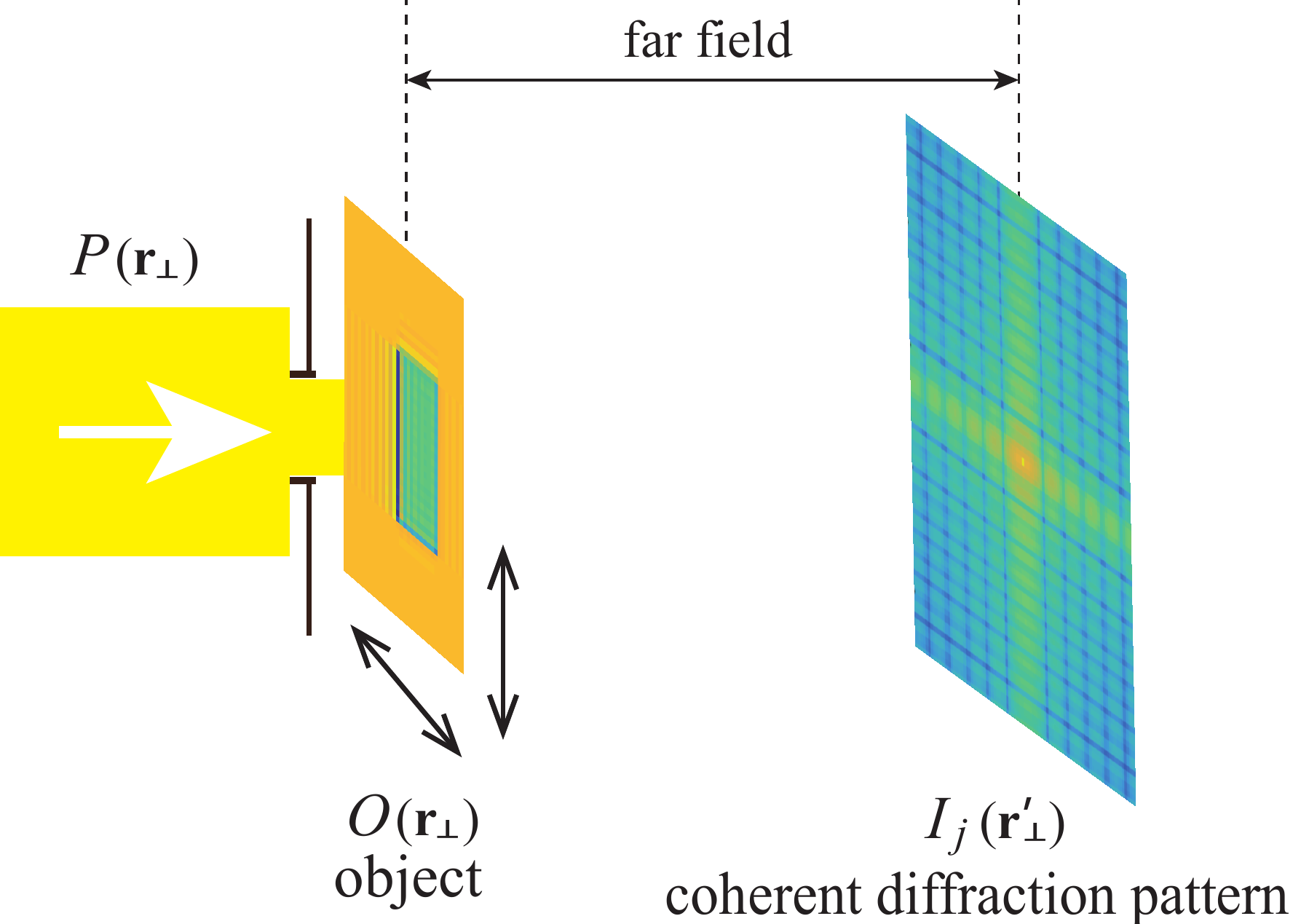}	
	\caption{The configuration of application 2.}\label{Fig.4-1}
\end{figure}

Now we consider a real-space ptychography setup as shown in Fig. \ref{Fig.4-1}. 
The object can be written in the following form:
\begin{align}
O(\textbf{r}_{\perp})(A_{1},\phi_{1},a_{1},b_{1},\textbf{r}_{\perp,1})\,&=\,1+(A_{1}e^{\text{i}\phi_{1}}-1)\Pi_{a_{1},b_{1}}(\textbf{r}_{\perp}-\textbf{r}_{\perp,1})\nonumber\\
&=\,1+C_{1}\Pi_{a_{1},b_{1},\textbf{r}_{\perp,1}}
\label{eq.4-3-1},
\end{align}
where $C_{1}=A_{1}e^{\text{i}\phi_{1}}-1$ is a complex valued coefficient. and $\Pi_{a_{1},b_{1}}(\textbf{r}_{\perp}-\textbf{r}_{\perp,1})$ is the 2D rectangular function defined by parameters:
\begin{align}
\Pi_{a_{1},b_{1}}(\textbf{r}_{\perp}-\textbf{r}_{\perp,1})\,=\,\Pi_{a_{1}}(x-x_{1})\Pi_{b_{1}}(y-y_{1})\,=\,
\left\lbrace 
\begin{tabular}{ll}
$0$,&$|x-x_{1}|>\frac{a_{1}}{2}$ or $|y-y_{1}|>\frac{b_{1}}{2}$,\\
$1$,&$|x-x_{1}|<\frac{a_{1}}{2}$ and $|y-y_{1}|<\frac{b_{1}}{2}$.
\end{tabular}
\right.
\label{eq.4-3-2}
\end{align}
We aim to retrieve the parameters: 
\begin{align}
\Theta\,&=\,\left[A_{1},\phi_{1},a_{1},b_{1},\textbf{r}_{\perp,1}\right]^{T},\quad
\text{where:}\quad
A_{1}\in(0,1],\; a_{1}>0,\; b_{1}>0. \label{eq.4-3-3}
\end{align}
The diffracted wavefield in the far field for the $j$th illumination is:
\begin{align}
\mathcal{F}\left(\varPsi_{j}\right)\left(\textbf{k}'_{\perp}\right)\,=\,&\mathcal{F}\left(P_{j}\right)\left(\textbf{k}'_{\perp}\right)+\mathcal{F}\left(P_{j}\right)\left(\textbf{k}'_{\perp}\right)\otimes\left[C_{1}a_{1}b_{1}\text{sinc}\left(\frac{a_{1}k_{x}}{2}\right)\text{sinc}\left(\frac{b_{1}k_{y}}{2}\right) e^{\text{i}\textbf{k}'_{\perp}\cdot\textbf{r}_{\perp,1}}\right],\label{eq.4-3-4}
\end{align}
where $\otimes$ denotes convolution. Note in Eq. (\ref{eq.4-3-2}) we leave the values of the function at $x=x_{1}\pm\frac{a_{1}}{2}$ and $y=y_{1}\pm\frac{b_{1}}{2}$ be undefined because these values cannot be retrieved under the projection approximation given by Eq. (\ref{eq.4-1-2}). We can see in Eq. (\ref{eq.4-3-4}) that the diffracted wavefield is not a function w.r.t. the value of $O(\textbf{r}_{\perp})$ at position $x=x_{1}\pm\frac{a_{1}}{2}$ and $y=y_{1}\pm\frac{b_{1}}{2}$. The validity of the projection approximation have been discussed in \cite{Rodenburg1992a,Thibault2008} and we assume in the paper that this approximation is valid.

\begin{comment}
\subsection{Multiple objects in constant surrounding}
For the case of multiple rectangular objects embedded in constant surrounding, the transmission function is denoted by:
\begin{align}
O\,=\,1+\sum_{i=1}^{N}(A_{1}e^{\text{i}\phi_{1}}-1)\Pi_{a_{1},b_{1}}(\textbf{r}-\textbf{r}_{1})\,=\,1+\sum_{i=1}^{N}C_{1}\Pi_{a_{1},b_{1},\textbf{r}_{1}}.
\label{eq.4-4}
\end{align}
where $i=1,2,\cdots,N$ is the index of rectangular object. 

For the time being we assume there is no overlap area between every rectangulars.
\end{comment}

\subsection{Retrieving the parameter of the rectangle}
We can see in Eq. (\ref{eq.4-3-4}) that, when we have exact knowledge of the probe, the diffraction pattern is a function of the parameters of the rectangle. This fact offers us the chance to retrieve the parameters given in Eq. (\ref{eq.4-3-3}) from the measurements $I_{j}(\textbf{k}'_{\perp})$ for all $j$. In this section we propose and validate a feasible method to retrieve the parameters from a ptychographic measurement.

The first step of the proposed method is to reconstruct the object function in real space, denoted by: $\hat{O}(\textbf{r}_{\perp})$, from $I_{j}(\textbf{k}'_{\perp})$ for all $j$. This can be done by applying the PIE\cite{Faulkner2004,Rodenburg2004} algorithm or other ptychography algorithms\cite{Thibault2009,Marchesini2013,Zhong2016,Odstrcil2018}. Note that the discretization of $\textbf{r}_{\perp}$ and $\textbf{k}'_{\perp}$ follows Eq. (\ref{eq.4-1-6}). For noisy measurements, one may use the Maximum Likelihood estimator (MLE) if one can find a dominant noise model\cite{Godard2012,Thibault2012}. For the case of Poisson noise, we can apply gradient descent methods\cite{Murray1982,Fletcher1988} to minimize the likelihood function $\mathcal{L}_{P}$ given by Eq. (S7) in the Supplement. Note that $\hat{O}(\textbf{r}_{\perp})$ can be obtained even if the probe function is unknown, which is due to the data redundancy of the ptychographic measurement.

Once the minimum of the likelihood function is found, we can compute the Fourier transform of the reconstructed object, denoted by $\mathcal{F}(\hat{O})(\textbf{k}_{\perp})$. The spacing of grid  $\textbf{r}_{\perp}$ and $\textbf{k}_{\perp}$ is given in Eq. (\ref{eq.4-1-8}). The parameter of the rectangle can be retrieved by minimizing a cost function $\mathcal{G}$ defined by:
\begin{align}
\mathcal{G}\,=\,\left\|\mathcal{F}\left(\hat{O}-1\right)-C_{1}a_{1}b_{1}\text{sinc}\left(\frac{a_{1}k_{x}}{2}\right)\text{sinc}\left(\frac{b_{1}k_{y}}{2}\right) e^{\text{i}\textbf{k}_{\perp}\cdot\textbf{r}_{\perp,1}}\right\|^{2},\label{eq.4-3-5}
\end{align}
where $\left\|\cdot\right\|^{2}$ denotes the $l_{2}$ norm. To give an example about the relation between $\mathcal{G}$ and the rectangle parameters, we show in Fig. \ref{Fig.4-3} the value of $\mathcal{G}$ as a function of $a_{1}$ and $x_{1}$. The configuration parameter of Fig. \ref{Fig.4-3} will be given later in Section 4.2. It is seen that $\mathcal{G}$ is convex in the neighborhood of the actual $a_{1}$ and $x_{1}$, which offers us the chance to retrieve the parameter by minimizing $\mathcal{G}$. In order to find the minimum of $\mathcal{G}$, it will be beneficial to start the algorithm from a point close to the actual value. This starting point can be determined from $\hat{O}$(\textbf{r}). 

In summary, our proposed method includes the following steps:
\begin{enumerate}[(1)]
	\item Use a ptychographic algorithm to retrieve the complex valued wavefield $\hat{O}(\textbf{r}_{\perp})$.
	\item Find the lower and upper bound of $\Theta$ from  $\hat{O}(\textbf{r}_{\perp})$. $\Theta$ is the parameter vector defined by Eq. (\ref{eq.4-3-3}). These bounds are denoted by: $\Theta_{l}$ and $\Theta_{u}$.	
	\item Solve the following problem:
	\begin{align}
	\arg\min_{\Theta}\hspace{0.5em} \mathcal{G},\hspace{1em}
	\mathrm{subject\, to}\hspace{0.5em}\Theta_{l}\leq\Theta\leq\Theta_{u}.\label{eq.4-3-6}
	\end{align}
\end{enumerate}

\subsection{Simulation}

\begin{figure}[htp!]
	\centering\includegraphics[width=0.8\textwidth]{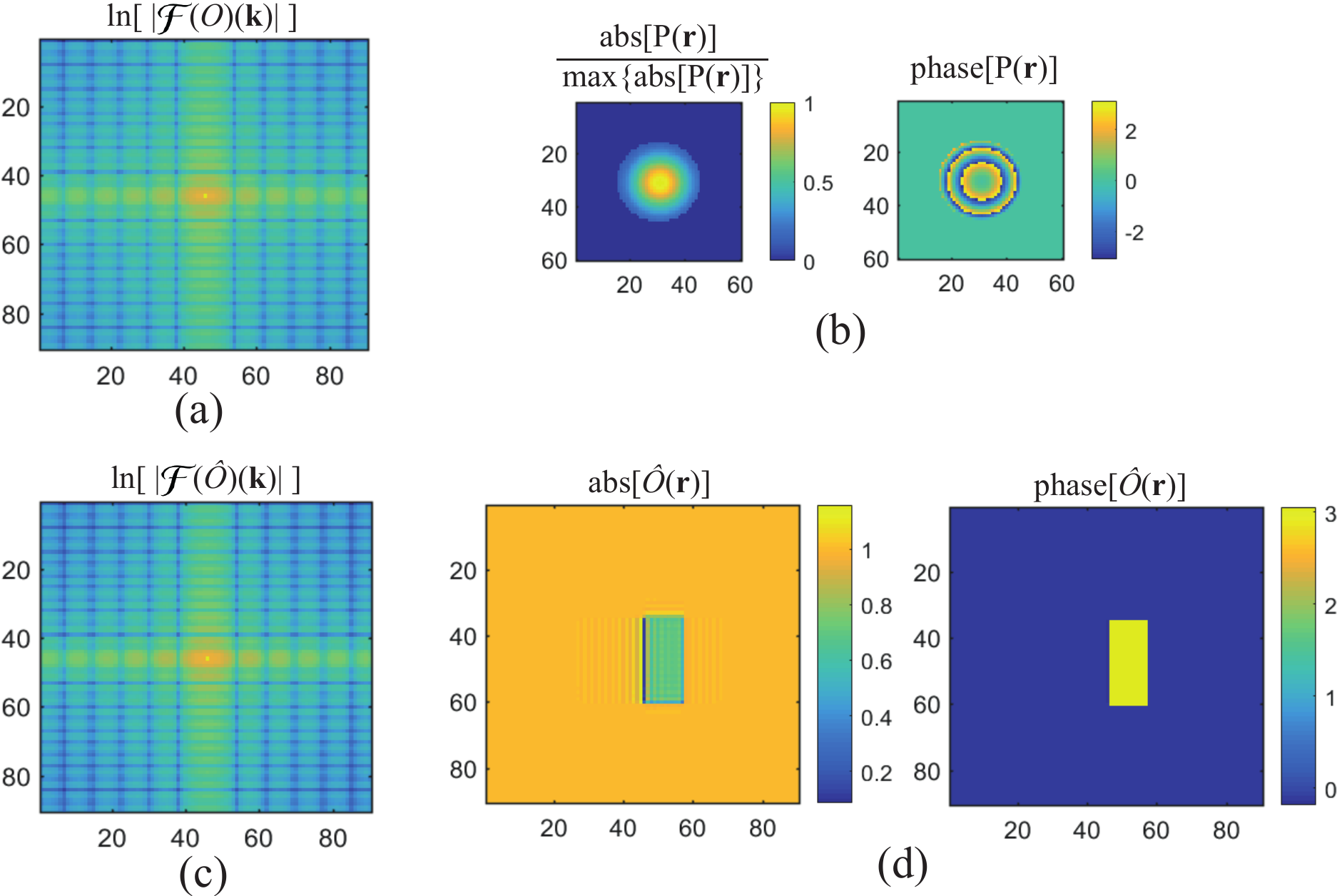}	
	\caption{(a) The simulated object in Fourier space. The object has one rectangle which is embedded in a constant surrounding. (b) The normalized amplitude and the phase of the probe, which is known in the simulation. (c) The retrieved object function in Fourier space from ptychographic measurement. (d) The inverse Fourier transform of $\mathcal{F}(\hat{O})(\textbf{k}_{\perp})$.}\label{Fig.4-2}
\end{figure}

To validate our proposed method, a preliminary simulation is shown. We consider the setup as shown in Fig. \ref{Fig.4-1}. Details of the configuration are shown in Table. \ref{table.4-1}. The Fresnel number of this configuration is 0.0014. According to Eq. (\ref{eq.4-3-4}), we first generate the complex valued wavefield in Fourier space $\mathcal{F}\left(\varPsi_{j}\right)\left(\textbf{k}'_{\perp}\right)$ based on the given probe and object. The Fourier transform of the object function $\mathcal{F}\left(O\right)\left(\textbf{k}_{\perp}\right)$ is illustrated in Fig. \ref{Fig.4-2}(a). The object consists of one rectangle with sizes listed in Table. \ref{table.4-2}. Fig. \ref{Fig.4-2}(b) shows the normalized amplitude and the phase of the probe. In this simulation we assume the probe is known and the ptychographic measurement is noise-free. In Fig. \ref{Fig.4-2}(c) we illustrate the Fourier transform of the reconstructed object function $\mathcal{F}(\hat{O})(\textbf{k}_{\perp})$. The inverse Fourier transform of $\mathcal{F}(\hat{O})(\textbf{k}_{\perp})$ is shown in Fig. \ref{Fig.4-2}(d).
\begin{table}
	\centering
	\caption{The characteristic parameters of the configuration in the simulation}
	\begin{tabular}{||c|c|c|p{0.1cm}|p{1.1cm}|c|c|c||}
		\hhline{|t:========:t|}
		\multirow{2}{*}{probe} & grid size & \makecell[c]{grid\\spacing} & \multicolumn{2}{c|}{wavelength} & \makecell[c]{scanning\\grid} & \makecell[c]{overlap\\ratio} & \makecell[c]{radius of\\circular support} \\
		\hhline{|~-------||}
		& $60\times60$ & $30nm$ & \multicolumn{2}{c|}{$30nm$} & $5\times5$ & $75\%$ & $0.45\mu m$\\
		\cline{4-4}\hhline{|:===~====:|}
		\multirow{2}{*}{object} & grid size & \makecell[c]{grid\\spacing} & & \multirow{2}{*}{detector} & \makecell[c]{pixel\\number}& pixel size & \makecell[c]{propagation\\distance} \\
		\hhline{|~--~~---||}
		& $90\times90$ & $30nm$ & & & $60\times60$ & $50\mu m$ & $1.88cm$ \\
		\hhline{|b:===b-b====:b|}
	\end{tabular}
	\label{table.4-1}
\end{table}

\begin{figure}[htp!]
	\centering\includegraphics[width=0.45\textwidth]{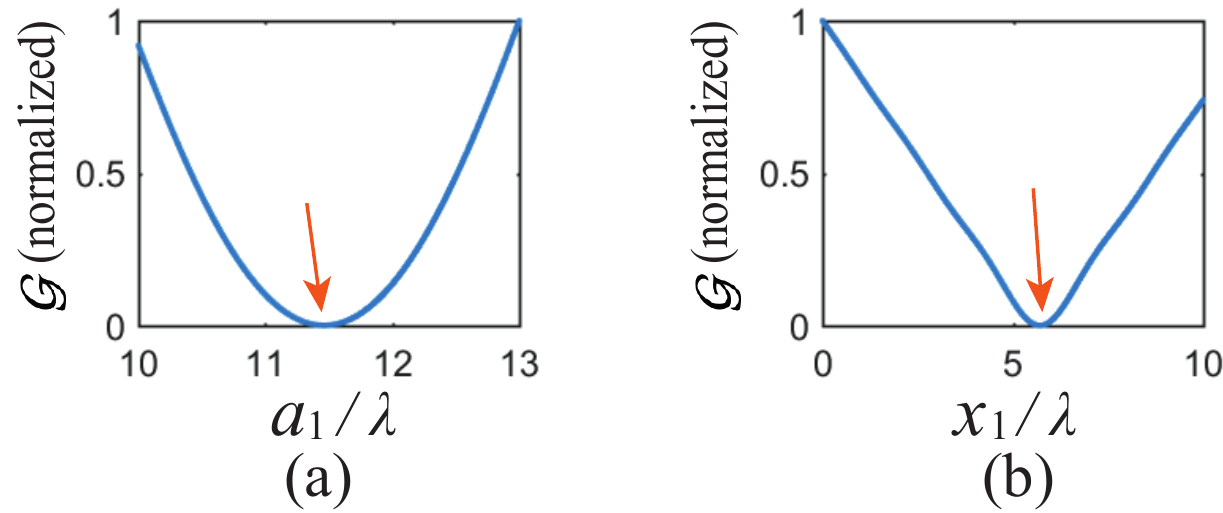}	
	\caption{The evaluation of $\mathcal{E}$ with respect to $a_{1}$ and $x_{1}$. The value of  $\mathcal{E}$ is normalized to its maximum in both plots. The orange arrow points to the actual value of $a_{1}$ and $x_{1}$ in this simulation.}\label{Fig.4-3}
\end{figure}

After obtaining $\mathcal{F}(\hat{O})(\textbf{k})$, we can retrieve the parameters of the rectangle by solving the optimization problem in Eq. (\ref{eq.4-3-6}). In Fig. \ref{Fig.4-3} we demonstrate the evaluation of the cost function $\mathcal{G}$ with respect to $a_{1}$ and $x_{1}$, which are the width and position of the rectangular in the
$x$-direction. The orange arrows in both plots points to the actual values of $a_{1}$ and $x_{1}$. We see in Fig. \ref{Fig.4-3} that it is possible to accurately retrieve the values of $a_{1}$ and $x_{1}$ by minimizing $\mathcal{G}$. To compute the solution of the problem in Eq. (\ref{eq.4-3-6}), we again implemented the 'fmincon' solver in MATLAB. Furthermore, Fig. \ref{Fig.4-3} shows that the value of $\mathcal{G}$ is approximately a quadratic function w.r.t. $a_{1}$ and a linear function w.r.t. $x_{1}$ in the neighborhood of the actual values, which is explained in Section 2 of Supplement. The actual value of the parameters, the starting point and the retrieved results are presented in Table. \ref{table.4-2}. We can see that the proposed method can successfully retrieve the parameters of the rectangle. 

\begin{table}
	\centering
	\caption{Retrieved parameters of one rectangle}
	\begin{tabular}{c||cccccc}
		\hline
		& $a_{1}/\lambda$ & $b_{1}/\lambda$ & $x_{1}/\lambda$ & $y_{1}/\lambda$ & $A_{1}$ & $\phi_{1}$ \\
		\hhline{=||======}
		actual value & $11.46$ & $25.99$ & $5.71$ & $1.42$ & $0.70$ & $3.14$ \\
		\hline
		initial guess & $11.00$ & $28.00$ & $4.00$ & $3.00$ & $0.73$ & $3.17$\\
		\hline
		retrieved value & $11.46$ & $25.99$ & $5.71$ & $1.42$ & $0.70$ & $3.14$\\
		\hline
	\end{tabular}
	\label{table.4-2}
\end{table}

\section{The CRLB analysis of the parameter retrieval scheme for Poisson noise}
In estimation theory, the Cram\'{e}r Rao Lower Bound (CRLB) gives a lower bound on the variance of any unbiased estimator for a parameter that is to be estimated. The estimators that can reach the lower bound are called the minimum variance unbiased estimators. Minimum variance unbiased estimators are often not available \cite{Kay2009,Bouchet2020}. To find the CRLB, one needs to compute the Fisher information matrix which is the expectation value of the second order derivative of the likelihood function. Detailed description about CRLB, Fisher information matrix and Maximum Likelihood Estimation is given in Section 1.A of Supplement.

In this paper we study the CRLB for Poisson distributed photon counting noise, which is the most dominant source of noise which naturally occurs even under the best experimental conditions\cite{Godard2012,Thibault2012}. The expectation of the second order derivative of the Poisson likelihood function can be found in Section 1.B of Supplement.

\section{The CRLB analysis of application 1}

\subsection{The Fisher information matrix for retrieval of the dipole}

Now we calculate the Fisher matrix for the $i$th dipole. According to Eq. (\ref{eq.4-2-12}), the parameters we aim to estimate are:
\begin{align}
\Theta\,=\,\left[\theta_{1},\theta_{2},\cdots,\theta_{N}\right]^{T}\,=\,\left[\alpha_{1},x_{1},y_{1},\alpha_{2},x_{2},y_{2},\cdots,\alpha_{N},x_{N},y_{N}\right]^{T}.\label{eq.4-5-1}
\end{align} 
We consider that we aim to retrieve the parameters of the $i$th dipole while assuming that the parameters of all other dipoles are known. To find the Fisher matrix, we need to calculate the derivative of $I_{j}$ with respect to the parameters of dipole $i$. The derivatives of $I_{j}$ are given in Section 1.C of the Supplement. The number of elements of $I_{j}$ are determined by the amount of dipoles.
For the case of two dipoles in application 1, we have the $6\times6$ Fisher matrix with elements:
\begin{flalign}
I_{F}^{\text{dip}}\,&=\,
\begin{bmatrix}
I_{F,\alpha_{1}\alpha_{1}}^{\text{dip}}&I_{F,\alpha_{1} \textbf{r}_{\perp,1}}^{\text{dip}}&I_{F,\alpha_{1}\alpha_{2}}^{\text{dip}}&I_{F,\alpha_{1} \textbf{r}_{\perp,2}}^{\text{dip}}\\
I_{F,\textbf{r}_{\perp,1}\alpha_{1}}^{\text{dip}}&I_{F,\textbf{r}_{\perp,1}\textbf{r}_{\perp,1}}^{\text{dip}}&I_{F,\textbf{r}_{\perp,1}\alpha_{2}}^{\text{dip}}&I_{F,\textbf{r}_{\perp,1}\textbf{r}_{\perp,2}}^{\text{dip}}\\
I_{F,\alpha_{2}\alpha_{1}}^{\text{dip}}&I_{F,\alpha_{2} \textbf{r}_{\perp,1}}^{\text{dip}}&I_{F,\alpha_{2}\alpha_{2}}^{\text{dip}}&I_{F,\alpha_{2} \textbf{r}_{\perp,2}}^{\text{dip}}\\
I_{F,\textbf{r}_{\perp,2}\alpha_{1}}^{\text{dip}}&I_{F,\textbf{r}_{\perp,2}\textbf{r}_{\perp,1}}^{\text{dip}}&I_{F,\textbf{r}_{\perp,2}\alpha_{2}}^{\text{dip}}&I_{F,\textbf{r}_{\perp,2}\textbf{r}_{\perp,2}}^{\text{dip}}
\end{bmatrix}
,\label{eq.4-5-4}
\end{flalign}
where $I_{F,\textbf{r}_{\perp,i}\textbf{r}_{\perp,i}}^{\text{dip}}$, $I_{F,\textbf{r}_{\perp,i}\alpha_{\perp,i}}^{\text{dip}}$ and
$I_{F,\alpha_{\perp,i}\textbf{r}_{\perp,i}}^{\text{dip}}$ are $2\times2$, $2\times1$ and $1\times2$ sub-matrices, respectively. The diagonal elements of $I_{F}^{\text{dip}}$ are:
\begin{flalign}
&I_{F,\alpha_{i}\alpha_{i}}^{\text{dip}}\,=\,\frac{2}{\hbar\omega}\sum_{\textbf{r}'_{\perp},j}\left[\frac{\left|\mathcal{F}^{-1}\left(\varPsi_{j,i}\right)(\textbf{r}'_{\perp})\right|^{2}}{\alpha_{i}^{2}}+\Re\left(\frac{\mathcal{F}^{-1}\left(\varPsi_{j}\right)^{*}(\textbf{r}'_{\perp})\left[\mathcal{F}^{-1}\left(\varPsi_{j,i}\right)(\textbf{r}'_{\perp})\right]^{2}}{\alpha_{i}^{2}\mathcal{F}^{-1}\left(\varPsi_{j}\right)(\textbf{r}'_{\perp})}\right)\right],\label{eq.4-5-5}\\
&I_{F,\textbf{r}_{\perp,i}\textbf{r}_{\perp,i}}^{\text{dip}}\,=\,\frac{2}{\hbar\omega}\sum_{\textbf{r}'_{\perp},j}\left[\left|\nabla_{\textbf{r}'_{\perp}}\mathcal{F}^{-1}\left(\varPsi_{j,i}\right)(\textbf{r}'_{\perp})\right|^{2}+\Re\left(\frac{\mathcal{F}^{-1}\left(\varPsi_{j}\right)^{*}(\textbf{r}'_{\perp})\left[\nabla_{\textbf{r}'_{\perp}}\mathcal{F}^{-1}\left(\varPsi_{j,i}\right)(\textbf{r}'_{\perp})\right]^{2}}{\mathcal{F}^{-1}\left(\varPsi_{j}\right)(\textbf{r}'_{\perp})}\right)\right],\label{eq.4-5-6}
\end{flalign}
which are given by Eq. (S17) and Eq. (S18) in the supplementary document.

\begin{comment}
where we used the following relation:
\begin{eqnarray}
\Re(z)\,=\,\frac{1}{2}\left[\Re(z)+\Re(z^{*})\right],\label{eq.4-5-7}
\end{eqnarray}
for any complex number $z$.	
\end{comment}

It is of interest to first study the diagonal terms in $I_{F}$. For instance, suppose that we have exact knowledge about the illumination power, the first dipole's position and the second dipole's strength and position, then $(I_{F,\alpha_{1}\alpha_{1}}^{\text{dip}})^{-1}$ is the CRLB of $\alpha_{1}$ for any unbiased estimator. 
When only one dipole exists in the sample, the diagonal terms in $I_{F}^{\text{dip}}$ can be rewritten as:
\begin{flalign}
&I_{F,\alpha_{1}\alpha_{1}}^{\text{dip}}\,=\,\frac{4}{\hbar\omega}\sum_{\textbf{r}'_{\perp},j}\left|\mathcal{F}^{-1}\left[Q(\textbf{k}_{\perp}+\textbf{k}_{\perp,j},z)e^{-\text{i}\textbf{k}_{\perp}\cdot\textbf{r}_{\perp,1}}\right]\right|^{2},\label{eq.4-5-8}\\
&I_{F,\textbf{r}_{\perp,1}\textbf{r}_{\perp,1}}^{\text{dip}}\,=\,\frac{4}{\hbar\omega}\left|C_{1}^{\text{dip}}\right|^{2}\sum_{\textbf{r}'_{\perp},j}\left[\frac{J_{2}\left(k\text{NA}\left|\textbf{r}'_{\perp}-\textbf{r}_{\perp,1}\right|\right)^{2}}{\left|\textbf{r}'_{\perp}-\textbf{r}_{\perp,1}\right|^{2}}\right],\label{eq.4-5-9}
\end{flalign}
where $C_{1}^{\text{dip}}$ is the complex valued constant:
\begin{align}
C_{1}^{\text{dip}}\,=\,\frac{\alpha_{1}A_{\text{in}}e^{\text{i}k_{z}\left|z\right|}k^{4}\text{NA}^{2}}{8\text{i}\pi\epsilon_{0}k_{z}}.\label{eq.4-5-10}
\end{align} 
In Eq. (\ref{eq.4-5-9}) we used the following relation\cite{GeorgeB.Arfken2012}:
\begin{align}
\frac{d}{dx}\left(\frac{J_{1}(x)}{x}\right)\,=\,\frac{-J_{2}(x)}{x},\label{eq.4-5-11}
\end{align} 
where $J_{1}$ and $J_{2}$ are the Bessel function of the first kind of order $1$ and $2$, respectively. 

We can see in Eq. (\ref{eq.4-5-8}) that the CRLB of $\alpha_{i}$ is inversely proportional to the total illumination power $A_{\text{in}}^{2}$. Therefore, it is needed to enhance the illumination power to determine the value of $\alpha_{i}$ for smaller particles. However,when the illumination power is enhanced too much, one may reach a saturation point due to the limited dynamic range of the detector. By taking dark field images of the sample, as shown in Fig. \ref{Fig.4-6}, one can avoid this limit. Furthermore, we observe that $I_{F,\textbf{r}_{\perp,1}\textbf{r}_{\perp,1}}^{\text{dip}}$ does not only depend on the values of $A$ and $\alpha_{1}$, but also on the NA. Therefore, to decrease the CRLB of $\textbf{r}_{\perp,1}$, one can increase the illumination power or one can enlarge NA, or enhance both. It is interesting that $I_{F,\textbf{r}_{\perp,1}\textbf{r}_{\perp,1}}^{\text{dip}}$ is not a function of $\textbf{k}_{\perp,j}$, which indicate that adjusting the illumination's incident angle can lead to any change of the CRLB of $\textbf{r}_{\perp,1}$ for the case of a single particle. 

When more than one particle is on the planar surface, we have to calculate the Fisher information by Eq. (\ref{eq.4-5-4}). We see from these equations that there is a correlation between the particles. Suppose there are two particles, then the CRLB of one of the particles is a function of the parameters of the other particle, as follows from Eq. (\ref{eq.4-5-5}) and Eq. (\ref{eq.4-5-6}) where the second terms consist of the complete field $\varPsi_{j}$ instead of only the partial field $\varPsi_{j,i}$. A more detailed study of the cross-correlation is presented in the next section.

\subsection{The CRLB of the dipole}
We study the CRLB of the dipole strength and the position of the dipole along the $x$-axis. We follow the configuration as described in Fig. (\ref{Fig.4-6}) and Table. \ref{table.4-3}. We first investigate the variance and the squared bias of parameters, $\alpha_{1}$ and $x_{1}$, of the dipole $i=1$. To find the variance and bias for various noise levels, we define the illumination power by counting the time-averaged number of photons scattered by the dipole $i=1$, which is given by:
\begin{align}
\text{PN}^{\text{dip}}\,=\,\frac{\left\| \mathcal{F}^{-1}\left(A_{\text{in}}k^{2}\frac{e^{\text{i}k_{z}\left|z\right|}}{8\text{i}\pi\epsilon_{0}k_{z}}\alpha_{i}e^{-\text{i}\textbf{r}_{\perp,i}\cdot\textbf{k}_{\perp}}\right)\right\|^{2}_{i=1}}{\hbar\omega}.\label{eq.4-5-12}
\end{align} 
The variance and bias are obtained from Monte Carlo simulations. We generated 1000 Fourier ptychographic dark field data-sets for $\text{PN}^{\text{dip}}=10^{4},10^{6},10^{8}$. The parameters are reconstructed from the data-sets by applying the parameter retrieval algorithm described in Section 2.4. The variance and squared bias for $\text{PN}^{\text{dip}}=10^{4},10^{6},10^{8},$ are shown in Table. \ref{table.4-5}.

\begin{figure}[htp!]
	\centering\includegraphics[width=0.8\textwidth]{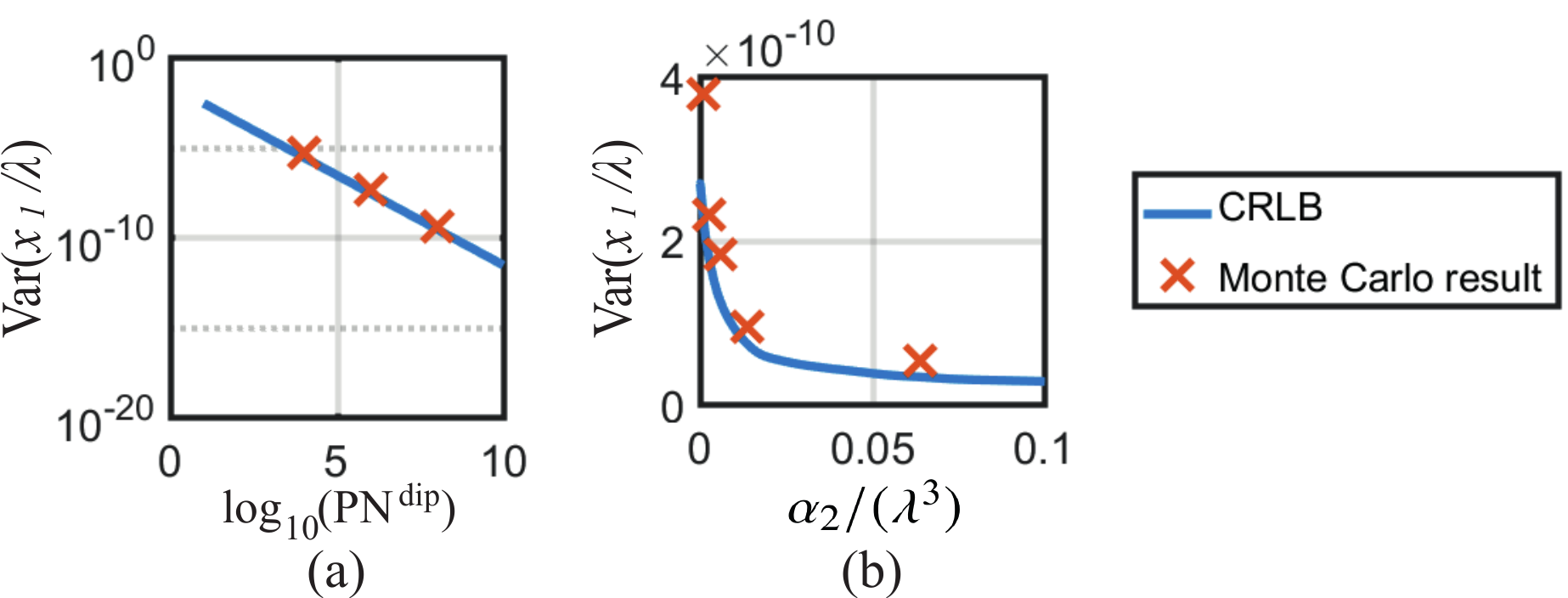}	
	\caption{(a) The computed CRLB and variance of the position of  the first dipole, i.e. $x_{1}$, for various  $\text{PN}^{\text{dip}}$. (b) The computed CRLB and variance of $x_{1}$ for various values of the polarisability of the second dipole, i.e. $\alpha_{2}$, for the case of $\text{PN}^{\text{dip}}=10^{8}$. The blue line of both plots are the computed CRLB and the red crosses show the variance obtained from the Monte Carlo experiment.}\label{Fig.4-8}
\end{figure}

When $\text{PN}^{\text{dip}}=10^{4}$, we see that the variance of $x_{1}$ obtained from the retrieval method is $10$ times larger than the squared bias. This variance-bias-ratio becomes higher when $\text{PN}^{\text{dip}}$ is increased. This observation means that the retrieval method of $x_{1}$ is asymptotically unbiased when $\text{PN}^{\text{dip}}>10^{4}$. These variances are illustrated in Fig. \ref{Fig.4-8}, together with the computed CRLB. It is shown that the variance of the retrieval of $x_{1}$ is indeed bounded by the CRLB when $\text{PN}^{\text{dip}}>10^{4}$. The value of the bound is inversely proportional to the value of $\text{PN}^{\text{dip}}$.
\begin{table}
	\centering
	\caption{The variance and squared bias of $\alpha_{1}$ and $x_{1}$ of the dipole of $i=1$ for different photon counts $\text{PN}^{\text{dip}}$, obtained from Monte Carlo result.}
	\begin{tabular}{c||ccc}
		\hline
		$\text{PN}^{\text{dip}}$ & $10^{4}$ & $10^{6}$ & $10^{8}$ \\
		\hhline{=||===}
		$\text{Var}\left[\alpha_{i=1}/(\lambda^{3})\right]$ & $3.14\times10^{-12}$ &  $2.54\times10^{-14}$ & $2.62\times10^{-16}$\\
		\hline
		$\text{Bias}\left[\alpha_{i=1}/(\lambda^{3})\right]^{2}$ & $3.22\times10^{-10}$ & $1.03\times10^{-13}$ & $1.26\times10^{-17}$\\
		\hline
	    $\text{Var}\left(x_{i=1}/\lambda\right)$ & $4.54\times10^{-6}$ & $4.28\times10^{-8}$ & $4.23\times10^{-10}$\\
		\hline
		$\text{Bias}\left(x_{i=1}/\lambda\right)^{2}$ & $4.32\times10^{-7}$ & $4.03\times10^{-11}$ & $1.89\times10^{-13}$\\
		\hline
	\end{tabular}
	\label{table.4-5}
\end{table}

However, Table. \ref{table.4-5} also shows that the variance of $\alpha_{1}$ obtained from the algorithm is much smaller than the squared bias when $\text{PN}^{\text{dip}}<10^{6}$, and indeed the retrieval algorithm of $\alpha_{1}$ is not unbiased as long as $\text{PN}^{\text{dip}}<10^{8}$ for the current setup. Therefore, the variance of the retrieved $\alpha_{i}$ may not be bounded by the CRLB when $\text{PN}^{\text{dip}}<10^{8}$. On the other hand, we can see in Eq. (\ref{eq.4-2-5}) that the accuracy of the reconstruction of $\alpha_{i}$ is not only influenced by the Poisson noise, but also by the fluctuation of the illumination power $A^{2}_{\text{in}}$. That is, the uncertainty about the exact value of $A$ will lead to uncertainty of the retrieval of $\alpha_{1}$. Therefore, it is more difficult to determine $\alpha_{1}$ than the position with the current scheme. 
 
\subsection{The correlation between two dipoles}
As has been noted in Section 5.1, when there are two particles on the surface, varying the parameters of one particle can lead to a change of the CRLB of the another particle. To verify this correlation between the particles, we calculated the CRLB of $x_{1}$ with various values of $\alpha_{2}$. The value of $\text{PN}^{\text{dip}}$ is chosen to be $10^{8}$ because the retrieval algorithm is asymptotically unbiased for this noise level, as has been shown in Section 5.2. The computed CRLB is validated by using Monte Carlo simulations, as illustrated in Fig. \ref{Fig.4-8}(b). 

It is seen in Fig. \ref{Fig.4-8}(b) that one can lower the CRLB of $x_{1}$ obtained from the algorithm by enhancing the scattering power of the dipole $i=2$. This observation can be understood by studying the property of the Poisson distribution. The signal-to-noise ratio (SNR) of Poisson noise is equal to $\sqrt{n(\textbf{r}'_{\perp})}$, where $n(\textbf{r}'_{\perp})$ is the number of photons detected by the pixel at $\textbf{r}'_{\perp}$. When the scattering power of particle $i=1$ is fixed, $n(\textbf{r}'_{\perp})$ is increased by enhancing the scattering power of the other particle, and therefore the signal-to-noise ratio of the system is increased. One may argue that this conclusion is inconsistent with the case where incoherent illumination is used. Let us imagine that we apply incoherent illumination to the setup in Fig. \ref{Fig.4-5}, then the radiation of each dipole is independent to the other and hence the image recorded by the detector is given by:
\begin{align}
I_{j}^{\text{incoh}}(\textbf{r}_{\perp}')\,=\,\sum_{i}\left|\mathcal{F}^{-1}\left[\alpha_{i}Q(\textbf{k}_{\perp}+\textbf{k}_{\perp,j})\right]\right|^{2}(\textbf{r}_{\perp}'-\textbf{r}_{\perp,i}')\,=\,\sum_{i}I_{j,i}^{\text{incoh}}(\textbf{r}_{\perp}').\label{eq.4-5-13}
\end{align}
When there are two dipoles, Eq. (\ref{eq.4-5-13}) shows that the signal of dipole $i=1$ is $I_{j,1}^{\text{incoh}}(\textbf{r}_{\perp}')$ whereas the variance of the signal is determined by $\sum_{i}I_{j,i}^{\text{incoh}}(\textbf{r}_{\perp}')$ at the neighborhood of position $\textbf{r}_{\perp,1}$. Therefore, for the case of incoherent illumination, the SNR of dipole $i=1$ should be decreased by enhancing the scattering power of the dipole $i=2$ because the variance is proportional to $\sum_{i}I_{j,i}^{\text{incoh}}(\textbf{r}_{\perp}')$ for Poisson noise. However, we emphasize that Eq. (\ref{eq.4-5-13}) is not the case of application 1. By comparing Eq. (\ref{eq.4-2-11}) to Eq. (\ref{eq.4-5-13}), we see that the measurement in application 1 contains the interference pattern of the point spread function of the dipoles. Hence, the conclusion for incoherent illumination is not applicable in application 1 and the SNR should be determined by the computed CRLB and the Monte Carlo simulations. Note that second order scattering is neglected in the current model, i.e. we ignore the scattered wavefield from the first particle which is excited by the second one because the particles are sparsely distributed on the sample.

\section{The CRLB analysis of application 2}

\subsection{Fisher information matrix for single rectangular object}
For application 2, the parameter vector we want to retrieve is:
\begin{align}
\Theta\,&=\,\left[\theta_{1},\theta_{2},\cdots\right]^{T}\,=\,\left[A_{1},\phi_{1},a_{1},b_{1},\textbf{r}_{\perp,1}\right]^{T},\label{eq.4-6-1}
\end{align}
To find the Fisher information matrix, we start from the expectation of the second order perturbation of $\mathcal{L}_{P}$:
\begin{flalign}
E\left(\delta^{2}{\mathcal{L}_{P}}\right)(\Theta)(\delta\Theta,\delta \tilde{\Theta})\,
&=\,\frac{2}{\hbar\omega}\sum_{\textbf{k}'_{\perp},j}\Re\left[\mathcal{F}\left[P_{j}\delta O(\Theta)(\delta\Theta)\right]\mathcal{F}\left[ P_{j}\delta O({\Theta})(\delta\tilde{\Theta})\right]^{*}\right]\nonumber\\
&\quad+\frac{2}{\hbar\omega}\sum_{\textbf{k}'_{\perp},j}\Re\left[\frac{\mathcal{F}\left( \varPsi_{j}\right)}{\mathcal{F}\left( \varPsi_{j}\right)^{*}}\mathcal{F}\left[P_{j}\delta O(\Theta)(\delta\Theta)\right]^{*}\mathcal{F}\left[ P_{j}\delta O({\Theta})(\delta\tilde{\Theta})\right]^{*}\right].\label{eq.4-6-2}
\end{flalign}
which is derived from Eq. (S11) in Supplement. The function $O$ is defined in Eq. (\ref{eq.4-3-1}). $\delta O$ is the derivative of $O$ w.r.t. $\Theta$. $\delta \Theta$ and $\delta \tilde{\Theta}$ are small perturbations of the parameters of the rectangle. The explicit expression of $\delta O$, $\delta \Theta$ and $\delta \tilde{\Theta}$ are given in Section 1.D of the Supplementary document.

By using Eq. (\ref{eq.4-6-2}), Eq. (S2) and Eq. (S23) in the Supplement, we obtain the diagonal elements of the Fisher matrix:
\begin{flalign}
I_{F,A_{1}A_{1}}^{\text{rect}}\,
%=\,&\frac{2}{\hbar\omega}\sum_{\textbf{k},j}\left|\mathcal{F}\left(P_{j}\Pi_{a_{1},b_{1},\textbf{r}_{1}}\right)\right|^{2}+\frac{2}{\hbar\omega}\sum_{\textbf{k},j}\Re\left[\frac{\mathcal{F}\left( \varPsi_{j}\right)}{\mathcal{F}\left( \varPsi_{j}\right)^{*}}e^{-2\text{i}\phi_{1}}\left[\mathcal{F}\left(P_{j}\Pi_{a_{1},b_{1},\textbf{r}_{1}}\right)^{*}\right]^{2}\right],\nonumber\\
=\,&\frac{2}{\hbar\omega}\sum_{\textbf{r},j}\left|P_{j}\Pi_{a_{1},b_{1},\textbf{r}_{1}}\right|^{2}+\frac{2}{\hbar\omega}\sum_{\textbf{r},j}\Re\left[\mathcal{F}^{-1}\left(\frac{\mathcal{F}\left( \varPsi_{j}\right)}{\mathcal{F}\left( \varPsi_{j}\right)^{*}}\right)e^{-2\text{i}\phi_{1}}\left[\left(P_{j}\Pi_{a_{1},b_{1},\textbf{r}_{1}}\right)^{*}\right]^{2}\right].\label{eq.4-IF_AiAi}
\end{flalign}
\begin{flalign}
I_{F,\phi_{1}\phi_{1}}^{\text{rect}}\,=\,&A_{1}^{2}I_{F,A_{1}A_{1}}.\label{eq.4-IF_PhiiPhii}
\end{flalign}
\begin{flalign}
I_{F,a_{1}a_{1}}^{\text{rect}}\,=\,&\frac{1}{2\hbar\omega}\sum_{y,j}\left|C_{1}\Pi_{b_{1},y_{1}}\right|^{2}\left[\left|P_{j}\right|^{2}(x_{1}+\frac{a_{1}}{2},y)+\left|P_{j}\right|^{2}(x_{1}-\frac{a_{1}}{2},y)\right]\nonumber\\
&+\frac{1}{2\hbar\omega}\sum_{y,j}\Re\left[\left(C_{1}^{*}\Pi_{b_{1},y_{1}}\right)^{2}\mathcal{F}^{-1}\left(\frac{\mathcal{F}\left(\varPsi_{j}\right)}{\mathcal{F}\left(\varPsi_{j}\right)^{*}}\right)(2x_{1}+a_{1},y)\left(P_{j}^{*}\right)^{2}(x_{1}+\frac{a_{1}}{2},y)\right]\nonumber\\
&+\frac{1}{2\hbar\omega}\sum_{y,j}\Re\left[\left(C_{1}^{*}\Pi_{b_{1},y_{1}}\right)^{2}\mathcal{F}^{-1}\left(\frac{\mathcal{F}\left(\varPsi_{j}\right)}{\mathcal{F}\left(\varPsi_{j}\right)^{*}}\right)(2x_{1}-a_{1},y)\left(P_{j}^{*}\right)^{2}(x_{1}-\frac{a_{1}}{2},y)\right]\nonumber\\\label{eq.4-IF_aiai}
&+\frac{1}{\hbar\omega}\sum_{y,j}\Re\left[\left(C_{1}^{*}\Pi_{b_{1},y_{1}}\right)^{2}\mathcal{F}^{-1}\left(\frac{\mathcal{F}\left(\varPsi_{j}\right)}{\mathcal{F}\left(\varPsi_{j}\right)^{*}}\right)(2x_{1},y)P_{j}^{*}(x_{1}+\frac{a_{1}}{2},y)P_{j}^{*}(x_{1}-\frac{a_{1}}{2},y)\right].
\end{flalign}
$I_{F,b_{1}b_{1}}^{\text{rect}}$ can be obtained by taking the above equation and interchanging $x$ with $y$ and $a_{1}$ with $b_{1}$.
\begin{flalign}
I_{F,x_{1}x_{1}}^{\text{rect}}\,=\,&\frac{2}{\hbar\omega}\sum_{y,j}\left|C_{1}\Pi_{b_{1},y_{1}}\right|^{2}\left[\left|P_{j}\right|^{2}(x_{1}+\frac{a_{1}}{2},y)+\left|P_{j}\right|^{2}(x_{1}-\frac{a_{1}}{2},y)\right]\nonumber\\
&+\frac{2}{\hbar\omega}\sum_{y,j}\Re\left[\left(C_{1}^{*}\Pi_{b_{1},y_{1}}\right)^{2}\mathcal{F}^{-1}\left(\frac{\mathcal{F}\left(\varPsi_{j}\right)}{\mathcal{F}\left(\varPsi_{j}\right)^{*}}\right)(2x_{1}+a_{1},y)\left(P_{j}^{*}\right)^{2}(x_{1}+\frac{a_{1}}{2},y)\right]\nonumber\\
&+\frac{2}{\hbar\omega}\sum_{y,j}\Re\left[\left(C_{1}^{*}\Pi_{b_{1},y_{1}}\right)^{2}\mathcal{F}^{-1}\left(\frac{\mathcal{F}\left(\varPsi_{j}\right)}{\mathcal{F}\left(\varPsi_{j}\right)^{*}}\right)(2x_{1}-a_{1},y)\left(P_{j}^{*}\right)^{2}(x_{1}-\frac{a_{1}}{2},y)\right]\nonumber\\\label{eq.4-IF_xixi}
&-\frac{4}{\hbar\omega}\sum_{y,j}\Re\left[\left(C_{1}^{*}\Pi_{b_{1},y_{1}}\right)^{2}\mathcal{F}^{-1}\left(\frac{\mathcal{F}\left(\varPsi_{j}\right)}{\mathcal{F}\left(\varPsi_{j}\right)^{*}}\right)(2x_{1},y)P_{j}^{*}(x_{1}+\frac{a_{1}}{2},y)P_{j}^{*}(x_{1}-\frac{a_{1}}{2},y)\right].
\end{flalign}
$I_{F,y_{1}y_{1}}^{\text{rect}}$ can be obtained by taking the above equation and interchanging $x$ with $y$ and $a_{1}$ with $b_{1}$.

We again focus on the diagonal elements of the Fisher matrix. Referring to the first term on the right-hand side of Eq. (\ref{eq.4-IF_AiAi}) and Eq. (\ref{eq.4-IF_PhiiPhii}), we can immediately see that the CRLB of $A_{1}$ and $\phi_{1}$ is partially determined by the illumination power. Similarly, in Eq. (\ref{eq.4-IF_aiai}) and Eq. (\ref{eq.4-IF_xixi}) we see that the CRLB of $a_{1}$ and $x_{1}$ is partially determined by the illumination power at $x_{1}\pm\frac{a_{1}}{2}$, which is the edge of the rectangle. We can also notice in Eq. (\ref{eq.4-IF_PhiiPhii}) that the CRLB of $\phi_{1}$ is inversely proportional to $A_{1}^{2}$. This observation means that one can retrieve $\phi_{1}$ more accurately by increasing the transmission of the rectangle, assuming that the estimator is unbiased. 

It is interesting that $I_{F,a_{1}a_{1}}^{\text{rect}}$ and $I_{F,x_{1}x_{1}}^{\text{rect}}$ are functions of $\Pi_{b_{1},y_{1}}$. This fact means that enlarging the width of the rectangle in the $y$-direction will decrease the CRLB of $a_{1}$ and $x_{1}$, which are parameters along the $x$-axis. This correlation between $b_{1}$ and the CRLB of $a_{1}$ and $x_{1}$ is demonstrated in the next subsection. The computed CRLB is validated by Monte Carlo simulations.

\subsection{The CRLB of the width and the position of the rectangle}
Now we consider the configuration of Section 3. As described in Section 5.2, we need to provide a measure of the noise level in terms of photon counting. For application 2, we define the illumination power by means of the total photon number counting over the cross section of the probe:
\begin{align}
\text{PN}^{\text{rect}}\,=\,\sum^{N_{x}^{\text{det}},N_{y}^{\text{det}}}_{\textbf{r}_{\perp}}\frac{\left\|P(\textbf{r}_{\perp})\right\|^2}{\hbar\omega},\label{eq.4-6-6}
\end{align} 
where the probe $P(\textbf{r}_{\perp})$ is shown in Fig. \ref{Fig.4-2}(b). 

\begin{figure}[htp!]
	\centering\includegraphics[width=1\textwidth]{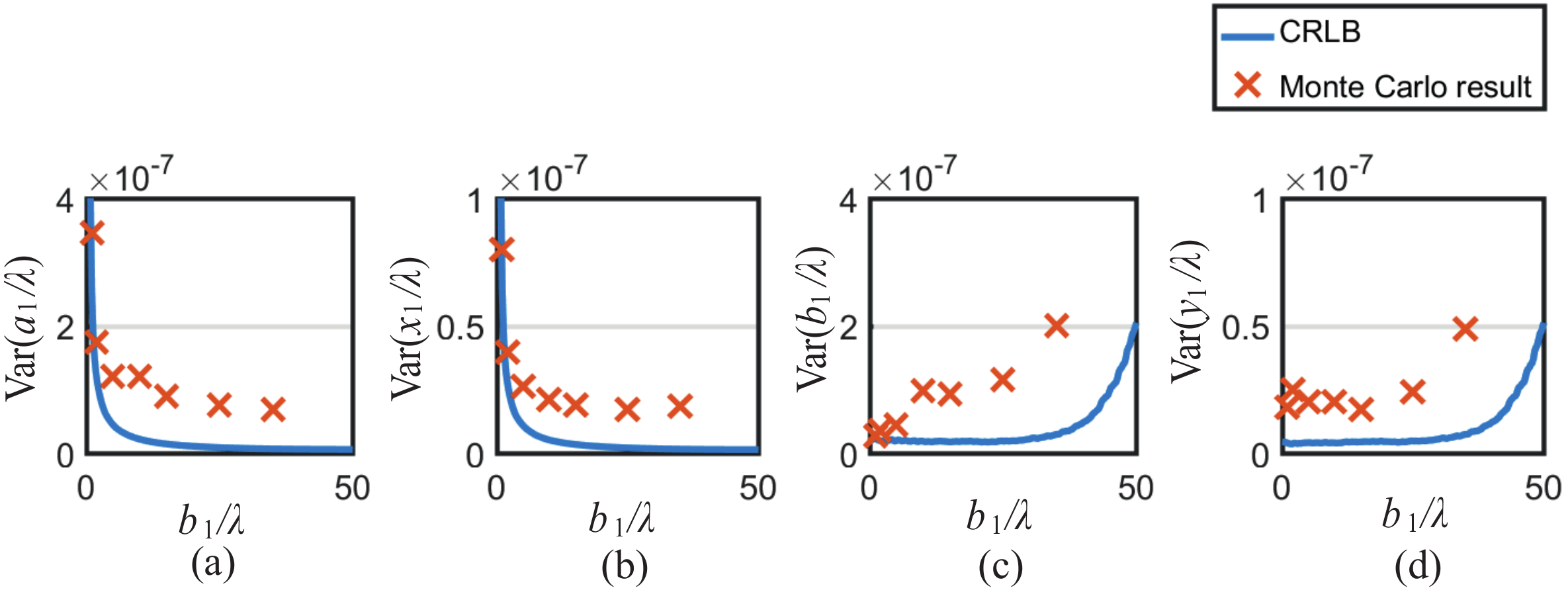}	
	\caption{The CRLB and variance of $a_{1}$, $x_{1}$, $b_{1}$ and $y_{1}$ of the rectangle, for various of $b_{1}$. The $\text{PN}^{\text{rect}}$ of this figure is $10^{8}$. The blue line is the computed CRLB and the red crosses show the variance obtained from the Monte Carlo simulations.}\label{Fig.4-10}
\end{figure}

Here we study the influence of the width of the rectangle in the $y$-direction on the variance of retrieved width and position along the $x$-axis. The computed CRLB of $a_{1}$ and $x_{1}$ are shown in Fig. \ref{Fig.4-10}, for various values of $b_{1}$. The value of $\text{PN}^{\text{rect}}$ is chosen to be $10^{8}$. To validate the computation of the CRLB, the result of Monte Carlo Monte simulations is shown in Fig. \ref{Fig.4-10} also. To obtain the variance, 1000 ptychographic data-sets are created in the Monte Carlo analysis. The data-sets are post-processed by using the parameter retrieval algorithm given in Section 3.2. The exact value of the variance and the squared bias of the parameters for the case of $b_{1}/\lambda=1,5,15,$ are listed in Table. \ref{table.4-6}.
\begin{table}
	\centering
	\caption{The variance and squared bias of $a_{1}$ and $x_{1}$ of the rectangle, obtained from the Monte Carlo simulation for $\text{PN}^{\text{rect}}=10^{8}$.}
	\begin{tabular}{c||ccc}
		\hline
		$b_{1}/\lambda$ & $1$ & $5$ & $15$  \\
		\hhline{=||===}
		$\text{Var}\left(a_{1}/\lambda\right)$ & $3.576\times10^{-7}$ & $1.455\times10^{-7}$ & $9.017\times10^{-8}$ \\
		\hline
		$\text{Bias}\left(a_{1}/\lambda\right)^{2}$ & $7.825\times10^{-10}$ & $4.254\times10^{-10}$ & $8.386\times10^{-11}$\\
		\hline
		$\text{Var}\left(x_{1}/\lambda\right)$ & $9.057\times10^{-8}$ & $2.527\times10^{-8}$ & $1.824\times10^{-8}$\\
		\hline
		$\text{Bias}\left(x_{1}/\lambda\right)^{2}$ & $6.423\times10^{-12}$ & $6.879\times10^{-11}$ & $4.947\times10^{-11}$\\
		\hline
	\end{tabular}
	\label{table.4-6}
\end{table}

We see in Fig. \ref{Fig.4-10}(a) and Fig. \ref{Fig.4-10}(b) that when $\text{PN}^{\text{rect}}=10^{8}$ the CRLB of $a_{1}/\lambda$ and $x_{1}/\lambda$ are in the order of $10^{-6}$, which indicates that the resolution of the current parameter retrieval scheme is not limited by the grid discretization in real space. The Monte Carlo result confirm this conclusion. Moreover, the squared bias of $a_{1}/\lambda$ and $x_{1}/\lambda$ is around $10^{3}$ times smaller that the variance, which means the that algorithm is asymptotically unbiased when $\text{PN}^{\text{rect}}=10^{8}$, and hence the variance obtained by the algorithm should be bounded by the CRLB.
The CRLB of both $a_{1}/\lambda$ and $x_{1}/\lambda$ decrease when the value of $b_{1}$ is increased. This result agrees with Eq. (\ref{eq.4-IF_aiai}) and Eq. (\ref{eq.4-IF_xixi}). The CRLB of $a_{1}/\lambda$ and $x_{1}/\lambda$ in Fig. \ref{Fig.4-10} decreases rapidly when $b_{1}/\lambda<5$. The reason is that the sensitivity of the retrieval of the parameters is determined by the number of photons which encodes the information about the parameters. That is, there are more photons which contain information about $a_{1}$ and $x_{1}$ when $b_{1}$ is larger. On the other hand, we can see that the CRLB of $b_{1}/\lambda$ and $y_{1}/\lambda$ do not vary much when the value of $b_{1}/\lambda$ is sufficiently small. When $b_{1}/\lambda>40$, the CRLB of $b_{1}/\lambda$ and $y_{1}/\lambda$ start to increase as the value of $b_{1}/\lambda$ is enlarged. This is because the boundary of the rectangle parallel to the $y$-axis falls outside of the illuminated area, which is an undesirable situation since $b_{1}/\lambda$ and $y_{1}/\lambda$ need also to be retrieved. To be explicit, the size of the illuminated area is determined by the non-zero area of $\sum_{j}\left|P(\textbf{r}_{\perp}-\textbf{R}_{\perp,j})\right|^{2}$. In the simulation, the size of the illuminated area in the $y$ direction is roughly $60\lambda$. Meanwhile, the beam profile of the illumination, i.e. $\left|P(\textbf{r}_{\perp})\right|^{2}$, is simulated by the Gaussian function as shown in Fig. \ref{Fig.4-2}(b). The full-width at half-maximum (FWHM) of the probe is around $15\lambda$. These characteristic parameters of the probe agree with Fig. \ref{Fig.4-10}(c) and Fig. \ref{Fig.4-10}(d) that the CRLB of $b_{1}/\lambda$ and $y_{1}/\lambda$ start to increase as $b_{1}/\lambda>40$. Overall, the computed CRLB as shown in Fig. \ref{Fig.4-10} indicates that, the optimal chosen range of values of $b_{1}/\lambda$ is $(5,40)$ for the current configuration.

%\section{Discussion}

\section{Conclusion}
In summary, a parameter retrieval method is demonstrated in this paper. The idea of the method is to incorporate available \textit{a priori} information about the object in the general ptychography framework. Two applications of the method are studied. In application 1 we explore how the parameters of small particles can be retrieved from Fourier ptychographic dark field measurements. The simulation result shows that, when the diameters of the particles are sufficiently small, e.g. $\sim 0.1\lambda$, so that the scattered wavefields can be modeled as radiation of dipoles, the parameters of the particles can be uniquely determined from dark field measurement only. In application 2 the retrieval of the parameters of a rectangular object embedded in constant surrounding was studied.

The influence of Poisson noise on the parameter retrieval method is discussed in the second part of the paper. The CRLB of the parameters are theoretically derived and numerically computed from the Fisher information matrix for both applications. Monte Carlo analysis is used to validate the computed CRLB. The CRLB, variance and bias of the retrieved parameters in application 1 were determined for various photon counts. It was found that the uncertainty of the parameter retrieval is inversely proportional to the photon counts, and potentially is not limited by the sizes of individual cells of the discretized meshgrid in object space. The correlation between at least two particles is evaluated from the calculation of the CRLB. We proved that the CRLB of the position of one particle is influenced by the scattering power of the other particle. This conclusion is confirmed by the Monte Carlo result. The correlation of parameters in application 2 is also inferred from the computed CRLB. The influence of the width of the rectangle in the $y$-direction on the CRLB of the parameters along the $x$-axis is investigated by analyzing the CRLB and the Monte Carlo result. For the same number of photons in the illuminating probe, the uncertainty of the parameters along the $x$-axis can be reduced by enlarging the width in the $y$-direction.

See \underline{Supplement 1} for supporting content.

\section*{Funding}
H2020 Marie Sk$\nmid$odowska-Curie Actions (675745).
	
\section*{Acknowledgments}	
X. Wei thanks O. el Gawhary for fruitful comments.

\section*{Disclosures} 
The authors declare no conflicts of interest.

%%%%%%%%%%%%%%%%%%%%%%% References %%%%%%%%%%%%%%%%%%%%%%%%%	
%%%%%%%%%% If using BibTeX:
\bibliography{sample}

\begin{thebibliography}{10}
\newcommand{\enquote}[1]{``#1''}

\bibitem{Hoppe1969}
W.~Hoppe, \enquote{Beugung im inhomogenen primärstrahlwellenfeld. i. prinzip
  einer phasenmessung von elektronenbeungungsinterferenzen,}
  {\protect\JournalTitle{Acta Crystallographica Section A}} \textbf{25},
  495--501 (1969).

\bibitem{Rodenburg1992a}
J.~M. Rodenburg and R.~H.~T. Bates, \enquote{The theory of super-resolution
  electron microscopy via wigner-distribution deconvolution,}
  {\protect\JournalTitle{Philosophical Transactions of the Royal Society of
  London. Series A: Physical and Engineering Sciences}} \textbf{339}, 521--553
  (1992).

\bibitem{Chapman1996}
H.~N. Chapman, \enquote{Phase-retrieval x-ray microscopy by wigner-distribution
  deconvolution,} {\protect\JournalTitle{Ultramicroscopy}} \textbf{66},
  153--172 (1996).

\bibitem{Faulkner2004}
H.~M.~L. Faulkner and J.~M. Rodenburg, \enquote{Movable aperture lensless
  transmission microscopy: A novel phase retrieval algorithm,}
  {\protect\JournalTitle{Physical Review Letters}} \textbf{93}, 023903 (2004).

\bibitem{Rodenburg2004}
J.~M. Rodenburg and H.~M.~L. Faulkner, \enquote{A phase retrieval algorithm for
  shifting illumination,} {\protect\JournalTitle{Applied Physics Letters}}
  \textbf{85}, 4795--4797 (2004).

\bibitem{Guizar-Sicairos2008}
M.~Guizar-Sicairos and J.~R. Fienup, \enquote{Phase retrieval with transverse
  translation diversity: a nonlinear optimization approach,}
  {\protect\JournalTitle{Optics Express}} \textbf{16}, 7264--7278 (2008).

\bibitem{Silva2015}
J.~C. da~Silva and A.~Menzel, \enquote{Elementary signals in ptychography,}
  {\protect\JournalTitle{Optics Express}} \textbf{23}, 33812--33821 (2015).

\bibitem{Seaberg2014}
M.~D. Seaberg, B.~Zhang, D.~F. Gardner, E.~R. Shanblatt, M.~M. Murnane, H.~C.
  Kapteyn, and D.~E. Adams, \enquote{Tabletop nanometer extreme ultraviolet
  imaging in an extended reflection mode using coherent fresnel ptychography,}
  {\protect\JournalTitle{Optica}} \textbf{1}, 39--44 (2014).

\bibitem{Odstrcil2015}
M.~Odstrcil, J.~Bussmann, D.~Rudolf, R.~Bresenitz, J.~Miao, W.~S. Brocklesby,
  and L.~Juschkin, \enquote{Ptychographic imaging with a compact
  gas{\textendash}discharge plasma extreme ultraviolet light source,}
  {\protect\JournalTitle{Optics Letters}} \textbf{40}, 5574--5577 (2015).

\bibitem{Rodenburg2007}
J.~M. Rodenburg, A.~C. Hurst, A.~G. Cullis, B.~R. Dobson, F.~Pfeiffer, O.~Bunk,
  C.~David, K.~Jefimovs, and I.~Johnson, \enquote{Hard-x-ray lensless imaging
  of extended objects,} {\protect\JournalTitle{Physical Review Letters}}
  \textbf{98}, 034801 (2007).

\bibitem{Thibault2008}
P.~Thibault, M.~Dierolf, A.~Menzel, O.~Bunk, C.~David, and F.~Pfeiffer,
  \enquote{High-resolution scanning x-ray diffraction microscopy,}
  {\protect\JournalTitle{Science}} \textbf{321}, 379--382 (2008).

\bibitem{Chapman2010}
H.~N. Chapman and K.~A. Nugent, \enquote{Coherent lensless x-ray imaging,}
  {\protect\JournalTitle{Nature Photonics}} \textbf{4}, 833--839 (2010).

\bibitem{Pfeiffer2017}
F.~Pfeiffer, \enquote{X-ray ptychography,} {\protect\JournalTitle{Nature
  Photonics}} \textbf{12}, 9--17 (2017).

\bibitem{Thibault2009}
P.~Thibault, M.~Dierolf, O.~Bunk, A.~Menzel, and F.~Pfeiffer, \enquote{Probe
  retrieval in ptychographic coherent diffractive imaging,}
  {\protect\JournalTitle{Ultramicroscopy}} \textbf{109}, 338--343 (2009).

\bibitem{Maiden2009}
A.~M. Maiden and J.~M. Rodenburg, \enquote{An improved ptychographical phase
  retrieval algorithm for diffractive imaging,}
  {\protect\JournalTitle{Ultramicroscopy}} \textbf{109}, 1256--1262 (2009).

\bibitem{Holler2017}
M.~Holler, M.~Guizar-Sicairos, E.~H.~R. Tsai, R.~Dinapoli, E.~Müller, O.~Bunk,
  J.~Raabe, and G.~Aeppli, \enquote{High-resolution non-destructive
  three-dimensional imaging of integrated circuits,}
  {\protect\JournalTitle{Nature}} \textbf{543}, 402--406 (2017).

\bibitem{Gardner2017}
D.~F. Gardner, M.~Tanksalvala, E.~R. Shanblatt, X.~Zhang, B.~R. Galloway, C.~L.
  Porter, R.~K. Jr, C.~Bevis, D.~E. Adams, H.~C. Kapteyn, M.~M. Murnane, and
  G.~F. Mancini, \enquote{Subwavelength coherent imaging of periodic samples
  using a 13.5{\hspace{0.25em}}nm tabletop high-harmonic light source,}
  {\protect\JournalTitle{Nature Photonics}} \textbf{11}, 259--263 (2017).

\bibitem{Jiang2018}
Y.~Jiang, Z.~Chen, Y.~Han, P.~Deb, H.~Gao, S.~Xie, P.~Purohit, M.~W. Tate,
  J.~Park, S.~M. Gruner, V.~Elser, and D.~A. Muller, \enquote{Electron
  ptychography of 2d materials to deep sub-{\aa}ngström resolution,}
  {\protect\JournalTitle{Nature}} \textbf{559}, 343--349 (2018).

\bibitem{Maiden2017}
A.~Maiden, D.~Johnson, and P.~Li, \enquote{Further improvements to the
  ptychographical iterative engine,} {\protect\JournalTitle{Optica}}
  \textbf{4}, 736--745 (2017).

\bibitem{Zheng2013}
G.~Zheng, R.~Horstmeyer, and C.~Yang, \enquote{Wide-field, high-resolution
  fourier ptychographic microscopy,} {\protect\JournalTitle{Nature Photonics}}
  \textbf{7}, 739--745 (2013).

\bibitem{Ou2014}
X.~Ou, G.~Zheng, and C.~Yang, \enquote{Embedded pupil function recovery for
  fourier ptychographic microscopy,} {\protect\JournalTitle{Optics Express}}
  \textbf{22}, 4960 (2014).

\bibitem{Horstmeyer2014}
R.~Horstmeyer and C.~Yang, \enquote{A phase space model of fourier
  ptychographic microscopy,} {\protect\JournalTitle{Optics Express}}
  \textbf{22}, 338 (2014).

\bibitem{Yeh2015}
L.-H. Yeh, J.~Dong, J.~Zhong, L.~Tian, M.~Chen, G.~Tang, M.~Soltanolkotabi, and
  L.~Waller, \enquote{Experimental robustness of fourier ptychography phase
  retrieval algorithms,} {\protect\JournalTitle{Optics Express}} \textbf{23},
  33214--33240 (2015).

\bibitem{Horstmeyer2015}
R.~Horstmeyer, R.~Y. Chen, X.~Ou, B.~Ames, J.~A. Tropp, and C.~Yang,
  \enquote{Solving ptychography with a convex relaxation,}
  {\protect\JournalTitle{New Journal of Physics}} \textbf{17}, 053044 (2015).

\bibitem{Zhang2015}
Y.~Zhang, W.~Jiang, and Q.~Dai, \enquote{Nonlinear optimization approach for
  fourier ptychographic microscopy,} {\protect\JournalTitle{Optics Express}}
  \textbf{23}, 33822 (2015).

\bibitem{Elser2003}
V.~Elser, \enquote{Phase retrieval by iterated projections,}
  {\protect\JournalTitle{Journal of the Optical Society of America A}}
  \textbf{20}, 40--55 (2003).

\bibitem{Godard2012}
P.~Godard, M.~Allain, V.~Chamard, and J.~Rodenburg, \enquote{Noise models for
  low counting rate coherent diffraction imaging,}
  {\protect\JournalTitle{Optics Express}} \textbf{20}, 25914--25934 (2012).

\bibitem{Thibault2012}
P.~Thibault and M.~Guizar-Sicairos, \enquote{Maximum-likelihood refinement for
  coherent diffractive imaging,} {\protect\JournalTitle{New Journal of
  Physics}} \textbf{14}, 063004 (2012).

\bibitem{Odstrcil2018}
M.~Odstr{\v{c}}il, A.~Menzel, and M.~Guizar-Sicairos, \enquote{Iterative
  least-squares solver for generalized maximum-likelihood ptychography,}
  {\protect\JournalTitle{Optics Express}} \textbf{26}, 3108--3123 (2018).

\bibitem{Maiden2011}
A.~M. Maiden, M.~J. Humphry, F.~Zhang, and J.~M. Rodenburg,
  \enquote{Superresolution imaging via ptychography,}
  {\protect\JournalTitle{Journal of the Optical Society of America A}}
  \textbf{28}, 604 (2011).

\bibitem{Szameit2012}
A.~Szameit, Y.~Shechtman, E.~Osherovich, E.~Bullkich, P.~Sidorenko, H.~Dana,
  S.~Steiner, E.~B. Kley, S.~Gazit, T.~Cohen-Hyams, S.~Shoham, M.~Zibulevsky,
  I.~Yavneh, Y.~C. Eldar, O.~Cohen, and M.~Segev, \enquote{Sparsity-based
  single-shot subwavelength coherent diffractive imaging,}
  {\protect\JournalTitle{Nature Materials}} \textbf{11}, 455--459 (2012).

\bibitem{Sidorenko2015}
P.~Sidorenko, O.~Kfir, Y.~Shechtman, A.~Fleischer, Y.~C. Eldar, M.~Segev, and
  O.~Cohen, \enquote{Sparsity-based super-resolved coherent diffraction imaging
  of one-dimensional objects,} {\protect\JournalTitle{Nature Communications}}
  \textbf{6} (2015).

\bibitem{Slepian1961}
D.~Slepian and H.~O. Pollak, \enquote{Prolate spheroidal wave functions,
  fourier analysis and uncertainty - i,} {\protect\JournalTitle{Bell System
  Technical Journal}} \textbf{40}, 43--63 (1961).

\bibitem{Papoulis1975}
A.~Papoulis, \enquote{A new algorithm in spectral analysis and band-limited
  extrapolation,} {\protect\JournalTitle{{IEEE} Transactions on Circuits and
  Systems}} \textbf{22}, 735--742 (1975).

\bibitem{Delsarte1985}
P.~Delsarte, A.~J. E.~M. Janssen, and L.~B. Vries, \enquote{Discrete prolate
  spheroidal wave functions and interpolation,} {\protect\JournalTitle{{SIAM}
  Journal on Applied Mathematics}} \textbf{45}, 641--650 (1985).

\bibitem{Walle2014}
P.~van~der Walle, S.~Hannemann, D.~van Eijk, W.~Mulckhuyse, and J.~C.~J.
  van~der Donck, \enquote{Implementation of background scattering variance
  reduction on the rapid nano particle scanner,} in \emph{Metrology,
  Inspection, and Process Control for Microlithography {XXVIII},}  J.~P. Cain
  and M.~I. Sanchez, eds. ({SPIE}, 2014).

\bibitem{Walle2017}
P.~van~der Walle, E.~Kramer, J.~C.~J. van~der Donck, W.~Mulckhuyse, L.~Nijsten,
  F.~A.~B. Arango, A.~de~Jong, E.~van Zeijl, H.~E.~T. Spruit, J.~H. van~den
  Berg, G.~Nanda, A.~K. van Langen-Suurling, P.~F.~A. Alkemade, S.~F. Pereira,
  and D.~J. Maas, \enquote{Deep sub-wavelength metrology for advanced defect
  classification,} in \emph{Optical Measurement Systems for Industrial
  Inspection X,}  P.~Lehmann, W.~Osten, and A.~A. Gon{\c{c}}alves, eds.
  ({SPIE}, 2017).

\bibitem{Ferreira1994}
P.~J.~S. Ferreira, \enquote{The stability of a procedure for the recovery of
  lost samples in band-limited signals,} {\protect\JournalTitle{Signal
  Processing}} \textbf{40}, 195--205 (1994).

\bibitem{Boef2016}
A.~J. den Boef, \enquote{Optical wafer metrology sensors for process-robust
  {CD} and overlay control in semiconductor device manufacturing,}
  {\protect\JournalTitle{Surface Topography: Metrology and Properties}}
  \textbf{4}, 023001 (2016).

\bibitem{Ansuinelli2019}
P.~Ansuinelli, W.~M.~J. Coene, and H.~P. Urbach, \enquote{Automatic feature
  selection in {EUV} scatterometry,} {\protect\JournalTitle{Applied Optics}}
  \textbf{58}, 5916 (2019).

\bibitem{Griffiths1999}
D.~J. Griffiths, \emph{Introduction to Electrodynamics (3rd Edition)} (Prentice
  Hall, 1999).

\bibitem{Novotny2012}
L.~Novotny and B.~Hecht, \emph{Principles of Nano-Optics} (Cambridge University
  Press, 2012).

\bibitem{Marchesini2013}
S.~Marchesini, A.~Schirotzek, C.~Yang, H.~tieng Wu, and F.~Maia,
  \enquote{Augmented projections for ptychographic imaging,}
  {\protect\JournalTitle{Inverse Problems}} \textbf{29}, 115009 (2013).

\bibitem{Zhong2016}
J.~Zhong, L.~Tian, P.~Varma, and L.~Waller, \enquote{Nonlinear optimization
  algorithm for partially coherent phase retrieval and source recovery,}
  {\protect\JournalTitle{{IEEE} Transactions on Computational Imaging}}
  \textbf{2}, 310--322 (2016).

\bibitem{Murray1982}
W.~Murray, M.~H. Wright, and P.~E. Gill, \emph{Practical Optimization} (Emerald
  Publishing Limited, 1982).

\bibitem{Fletcher1988}
R.~Fletcher, \emph{Practical Methods of Optimization, 2nd Edition} (Wiley,
  1988).

\bibitem{Kay2009}
S.~M. Kay, \emph{Fundamentals Of Statistical Signal Processing, Volume 1:
  Estimation Theory} (Pearson, 2009).

\bibitem{Bouchet2020}
D.~Bouchet, R.~Carminati, and A.~P. Mosk, \enquote{Influence of the local
  scattering environment on the localization precision of single particles,}
  {\protect\JournalTitle{Physical Review Letters}} \textbf{124} (2020).

\bibitem{GeorgeB.Arfken2012}
H.~J.~W. George B.~Arfken, \emph{Mathematical Methods for Physicists} (Elsevier
  LTD, Oxford, 2012).

\end{thebibliography}
	
\end{document}